\documentclass[a4paper,11pt]{article}
\usepackage{amsmath, amssymb, amsfonts,indentfirst,graphicx,color,anysize,bm}
\usepackage{fontenc,inputenc}
\usepackage{url}
\usepackage{xcolor}
\definecolor{myblue}{rgb}{0.0, 0.0, 0.6}
\usepackage{hyperref}
\hypersetup {
  colorlinks = true,
  citecolor  = myblue,
  linkcolor  = myblue,
  urlcolor   = myblue
}
\marginsize{3cm}{3cm}{2cm}{2cm}
\title{\includegraphics[width=0.35\textwidth]{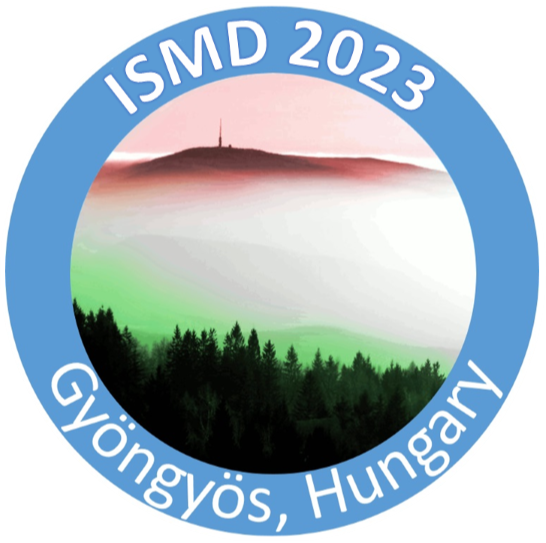}\\[0.7cm]
\textbf{\boldmath Polarization Measurements of $p^{\uparrow}$ and $^3$He$^\uparrow$ Beams at RHIC and Future EIC Using the Polarized Atomic Hydrogen Gas Jet Target}\thanks{Submitted to the Proceedings of the 52nd International Symposium on Multiparticle Dynamics, August 21--26, 2023, Gy\"ongy\"os, Hungary.} }
\author{{ A. A. Poblaguev}\\[1ex]
	Brookhaven National Laboratory, Upton, NY 11967, USA\\}
\date{January 11, 2024}

\begin{document}
\maketitle

\begin{abstract}
At the Relativistic Heavy Ion Collider (RHIC), the Polarized Atomic Hydrogen Gas Jet Target polarimeter (HJET) is employed for the precise measurement of the absolute transverse (vertical) polarization of proton beams, achieving low systematic uncertainties of approximately $\sigma^\text{syst}_P/P\leq0.5\%$. The acquired experimental data not only facilitated the determination of single $A_\text{N}(t)$ and double $A_\text{NN}(t)$ spin analyzing powers for 100 and 255 GeV proton beams, but also revealed a non-zero Pomeron spin-flip contribution through a Regge fit. Preliminary results obtained for forward inelastic $p^{\uparrow}p$ and elastic $p^{\uparrow}A$ analyzing powers will be discussed. The success of the HJET at RHIC suggests its potential application for proton beam polarimetry at the upcoming Electron--Ion Collider (EIC), aiming for an accuracy of 1\%. Moreover, the provided analysis indicates that the RHIC HJET target can serve as a tool for the precision calibration, with the required accuracy, of the $^3$He beam polarization at the EIC.
\end{abstract}

\pagestyle{plain}

\section{Introduction}

The basic requirements for beam polarimetry at the future Electron--Ion Collider \cite{Accardi:2012qut} include (i) non-destructive operation with minimal impact on the beam lifetime and (ii) low systematic uncertainty, as expressed by the condition
\begin{equation}
\sigma^\text{syst}_P / P \lesssim 1\%,
\label{eq:EIC_syst}
\end{equation}
in the value of the beam polarization \cite{AbdulKhalek:2021gbh}. This paper will focus on the absolute calibration of the EIC hadron beam polarization.

Hadron polarimetry has been successfully performed on the polarized proton beams at the Relativistic Heavy Ion Collider (RHIC) for nearly two decades. The absolute polarization of the proton beam (averaged over the beam profile) is measured using the Polarized Atomic Hydrogen Gas Jet Target polarimeter (HJET) \cite{Zelenski:2005mz}. This is achieved by experimentally determining the beam and jet spin asymmetries $a\!=\!A_\text{N}P$ for low-energy ($|t|<0.02$ GeV$^2$) recoil protons. The beam polarization is then related, using
\begin{equation}
  P_\text{beam}=P_\text{jet}\,a_\text{beam}/a_\text{jet},
  \label{eq:Pbeam}
\end{equation}
to the jet polarization $P_\text{jet}\approx0.96$, which is monitored with an accuracy of about $0.001$ by a Breit--Rabi polarimeter. For elastic scattering of the proton beam off the jet target proton, the result is independent of the details of the elastic $p^{\uparrow}p$ transverse analyzing power $A_\text{N}(s,t)$, which generally depends on the center-of-mass energy squared $s$ and momentum transfer squared $t$.

After an upgrade in 2015 \cite{Poblaguev:2016CD}, the HJET polarimeter provided the absolute beam profile-averaged polarization with systematic uncertainties of $\sigma^\text{syst}_P/P\!\lesssim\!0.5\%$ \cite{Poblaguev:2020qbw}. Additionally, HJET conducted measurements of single $A_\text{N}(t)$ and double $A_\text{NN}(t)$ spin-analyzing powers for 100 and 255 GeV proton beams. A Regge fit indicated a non-zero Pomeron spin-flip contribution. Preliminary results for forward inelastic $p^{\uparrow}p$ and elastic $p^{\uparrow}A$ (for six nuclei in the mass range from deuteron to gold) analyzing powers were also obtained. The effective operation of the HJET at RHIC suggests its potential application for proton beam polarimetry at the Electron--Ion Collider (EIC) with a targeted systematic accuracy of 1\%.

Furthermore, there has been a proposal\,\cite{Poblaguev:2022hqh} to utilize the HJET target for measuring the $^3$He ($h$) beam polarization at the EIC. It is recognized that the ratio $a_\text{beam}/a_\text{jet}$ used for calculating the beam polarization needs adjustment based on the ratio of $p^{\uparrow}h$ and $h^{\uparrow}p$ analyzing powers. In the leading-order Coulomb-nuclear interference (CNI) approximation, this ratio $A_\text{N}^{ph}/A_\text{N}^{hp}\!=\!-1.283$ is determined solely by the magnetic moments of the proton and helion \cite{Buttimore:2009zz}. However, it is important to note that considerations for hadronic spin-flip amplitudes and the potential breakup of $^3$He in the scattering process are necessary.

Based on the Glauber approach principles, it has been demonstrated that the $p^{\uparrow}h$ and $h^{\uparrow}p$ hadronic spin-flip amplitudes can be related with sufficient accuracy to the $p^{\uparrow}p$ amplitude measured at the HJET. The analysis of the breakup effect shows that, while it can introduce corrections to the spin-flip interference terms of up to 4\%, the overall effect cancels out to a negligible value in the analyzing power ratio. Detailed explanations of these estimates will be provided.

\section {The Polarized Atomic Hydrogen Gas Jet Target}

\begin{figure}[t]
  \begin{minipage}[c]{0.6\columnwidth}
    \begin{center}
      \includegraphics[width=1.0\columnwidth]{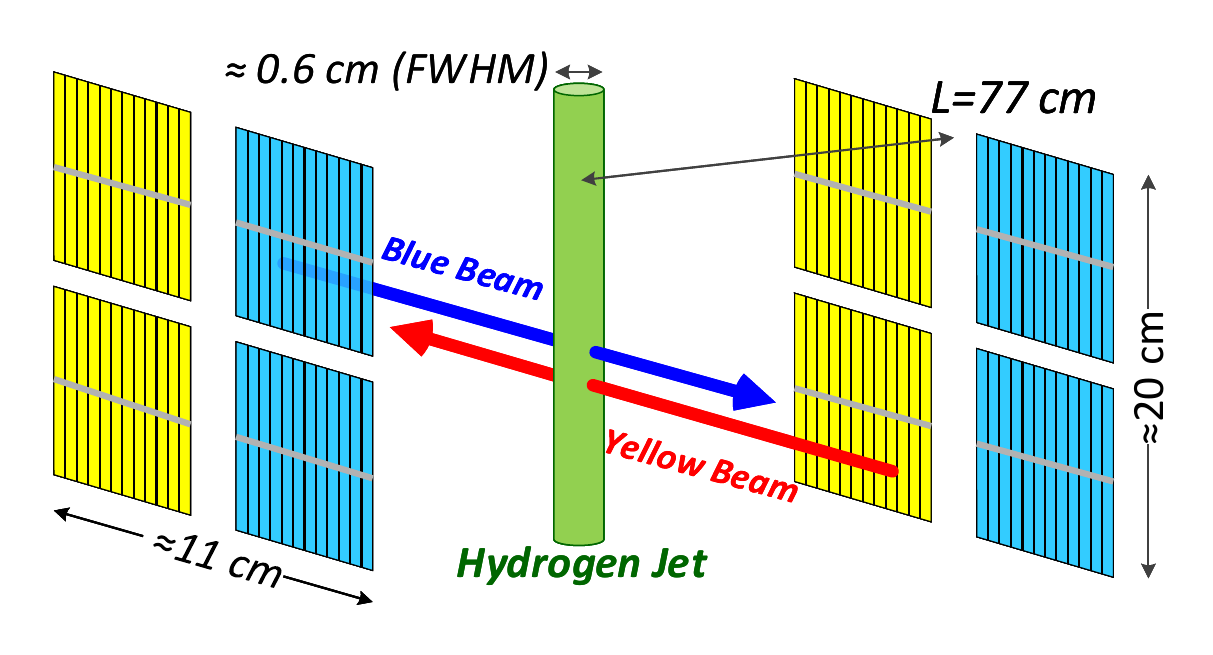}
    \end{center}
  \end{minipage}
  \hfill
  \begin{minipage}[c]{0.38\columnwidth}  
    \caption{ A schematic view of the HJET recoil spectrometer.
      \label{fig:HjetView}
    }
  \end{minipage}
\end{figure}

At the HJET (Fig.\,\ref{fig:HjetView}), vertically polarized proton beams scatter off the vertically polarized gas jet target, and the recoil protons are detected in the left/right symmetric, vertically oriented Si strips. These measurements are carried out continuously during the RHIC run and concurrently for the \textit{blue} and \textit{yellow} beams.

For the detected events, whether elastic or inelastic, the momentum transfer squared $t$ can be straightforwardly expressed in terms of the recoil proton kinetic energy $T_R$ and its mass $m_p$ as
\begin{equation}
  -t = 2m_pT_R,
\end{equation}
and it is constrained by the following CNI range defined by the Si detector geometry:
\begin{equation}
  0.0013 < -t < 0.018\,\text{GeV}^2.
\end{equation}

To confirm that the detected particle is a proton, the measured time of flight is matched with that derived from the measured $T_R$. For elastic events, the $z$-coordinate (along the beam) of the recoil proton in the detector, $z_R$, is given by
\begin{equation}
  \frac{z_R - z_\text{jet}}{L} = \sqrt{\frac{T_R}{2m_p}\times\frac{E_\text{beam} + m_p}{E_\text{beam} - m_p + T_R}},
  \label{eq:zR_el}
\end{equation}
where $z_\text{jet}$ is the coordinate of the scattering point, $L$ is the distance from the scattering point to the detector, and $E_\text{beam}$ is the beam energy. This equation allows for the isolation of elastic events. The remaining background, which is of only a few percent after applying the event selection cuts, can be reduced to a negligible level, as described in Ref.\,\cite{Poblaguev:2020qbw}.

The beam and jet spin asymmetries can be derived from the counted numbers of events in Si detectors\,\cite{Poblaguev:2020qbw} as functions of the recoil proton energy $T_R$ (or, equivalently, the momentum transfer squared $t$), and they can be related to the beam and jet polarizations as follows:
\begin{equation}
  a_\text{beam}(T_R) = A_\text{N}(s,t)\times P_\text{beam}, \qquad 
  a_\text{jet}(T_R)  = A_\text{N}(s,t)\times P_\text{jet}. 
\end{equation}

As the same events are used for calculating both the beam and jet asymmetries, the values $\langle a_\text{beam}(T_R)\rangle$ and $\langle a_\text{jet}(T_R)\rangle$, averaged over the entire range of $T_R$, can be substituted into Eq.\,(\ref{eq:Pbeam}) to determine the beam polarization. In this scenario, no prior knowledge of $A_\text{N}(s,t)$ is necessary. 

However, the analyzing power $A_\text{N}(s,t)$ for $s\!=\!2m_p(E_\text{beam}\!+\!m_p)$ can still be precisely derived from the measured $a_\text{jet}(T_R)$.

\section{\boldmath Single Spin Analyzing Power ${A_\text{N}(s,t)}$}

Omitting some small terms, the high-energy forward elastic $p^{\uparrow}p$ analyzing power can be expressed\,\cite{Kopeliovich:1974ee,Buttimore:1978ry,Buttimore:1998rj} via single spin-flip $\phi_5(t)$ and non-flip $\phi_+$ helicity amplitudes as
\begin{equation}
  A_\text{N}(t) = \frac{-2\mathrm{Im}\left(\phi_5^*\phi_+\right)}%
                      {\left|\phi_+\right|^2}%
  = \frac{2\mathrm{Im}\left[
      \phi_5^\text{em}{\phi_+^\text{h}}^* +
      \phi_5^\text{h}{\phi_+^\text{em}}^* +
      \phi_5^\text{h}{\phi_+^\text{h}}^*
      \right]}
  {\left| \phi_+^\text{h} + \phi_+^\text{em} e^{i\delta_C} \right|^2},
  \label{eq:AN_phi}
\end{equation}
where $\phi^\text{h}$ and $\phi^\text{em}$ are the hadronic and electromagnetic parts of the amplitudes, respectively, and $\delta_C$ is a Coulomb phase\,\cite{Cahn:1982nr,Kopeliovich:2000ez}. In Ref.\,\cite{Buttimore:1998rj}, $A_\text{N}(t)$ was rewritten in a form convenient for the experimental data analysis as
\begin{equation}
  A_\text{N}(t) = \frac{\sqrt{-t}}{m_p}\times\frac%
    {\kappa_p(1-\rho\delta_C)\,t_c/t~-~2(I_5-R_5\delta_C)\,t_c/t%
      ~-~2(R_5-\rho I_5) }%
    { (t_c/t)^2 - 2(\rho+\delta_C)\,t_c/t+1+\rho^2 },
    \label{eq:A_N}
\end{equation}
where $\kappa_p\!=\!1.793$ is the anomalous magnetic moment of the proton, $\rho\!=\!\mathrm{Re\!\;}\phi_+^\text{h}(t)/\mathrm{Im\!\;}\phi_+^\text{h}(0)$, and $R_5$ and $I_5$ are the real and imaginary parts of the hadronic spin-flip amplitude parameter\,\cite{Buttimore:1998rj}
\begin{equation}
  r_5=\frac{m_p\,\phi_5^\text{h}(t)}{\sqrt{-t}\,\mathrm{Im\!\;}\,\phi_+^\text{h}(0)}%
  = R_5+iI_5.
  \label{eq:r5_def}
\end{equation}
The electromagnetic amplitudes can be easily identified by the $t_c/t$ factor where, following the optical theorem, $t_c$ is related to the total cross-section as $-t_c\!=\!8\pi\alpha/\sigma_\text{tot}$. For a 100\,GeV proton beam, $\rho\!=\!-0.079$, $-t_c\!=\!1.86\!\times\!10^{-3}\,\text{GeV}^2$, $\delta_C\!=\!0.024\!+\alpha\!\ln(t_c/t)$. The precision measurement of $A_\text{N}(t)$ may allow one to experimentally determine the hadronic spin-flip amplitude. At the HJET, it was found that\,\cite{Poblaguev:2019saw} $|r_5|\sim0.02$ at RHIC energies (see Eqs.\,(\ref{eq:R5_100})--(\ref{eq:I5_255})).

Eq.\,(\ref{eq:A_N}) was standardly used in the experimental data analysis\,\cite{Okada:2005gu,Alekseev:2009zza,STAR:2012fiw}. However, it was noted\,\cite{Krelina:2019mlu,Kopeliovich:2021rdd} that, for already achieved experimental accuracy, neglecting the difference between hadronic and electromagnetic form factors in Eq.\,(\ref{eq:A_N}) may lead to the misinterpretation of the experimental results. A possible alteration of STAR\,\cite{STAR:2012fiw} and recent HJET\,\cite{Poblaguev:2019saw} values of $r_5$ are due to this,
\begin{equation}
  t_c/t \to t_c/t + r_E^2/3-B/2,
  \label{eq:ANcorr}
\end{equation}
and some other small corrections were evaluated in Ref.\,\cite{Poblaguev:2019vho}. $B\!=\!11.2\,\text{GeV}^2$\,\cite{Bartenev:1973jz} (for the 100\,GeV beam) is the non-flip differential cross-section slope and $r_E\!=\!0.841\,\text{fm}$\,\cite{Workman:2022ynf} is the rms charge radius of the proton.

It was also underlined in Refs.\,\cite{Krelina:2019mlu,Kopeliovich:2021rdd} that, due to the absorption, the effective charge radius of the proton in high energy $pp$ scattering differs from that\,\cite{Workman:2022ynf} found in electron--proton scattering and other lepton--proton interactions. In Eq.\,(\ref{eq:A_N}), the absorption corrections can be accounted for\,\cite{Kopeliovich:2021rdd,Poblaguev:2021xkd} by replacing
\begin{equation}
  R_5 \quad\to\quad R_5 - \frac{\alpha\kappa_p}{2}\,\frac{B}{B+2r_M^2/3} \quad\approx\quad R_5-0.003,
  \label{eq:R5_abs}
\end{equation}
  where $r_M\!=\!0.851\!\pm\!0.026\,\text{fm}$\,\cite{Lee:2015jqa} is the rms magnetic radius of the proton.

\begin{figure}[t]
  \begin{center}
    \includegraphics[width=0.45\columnwidth]{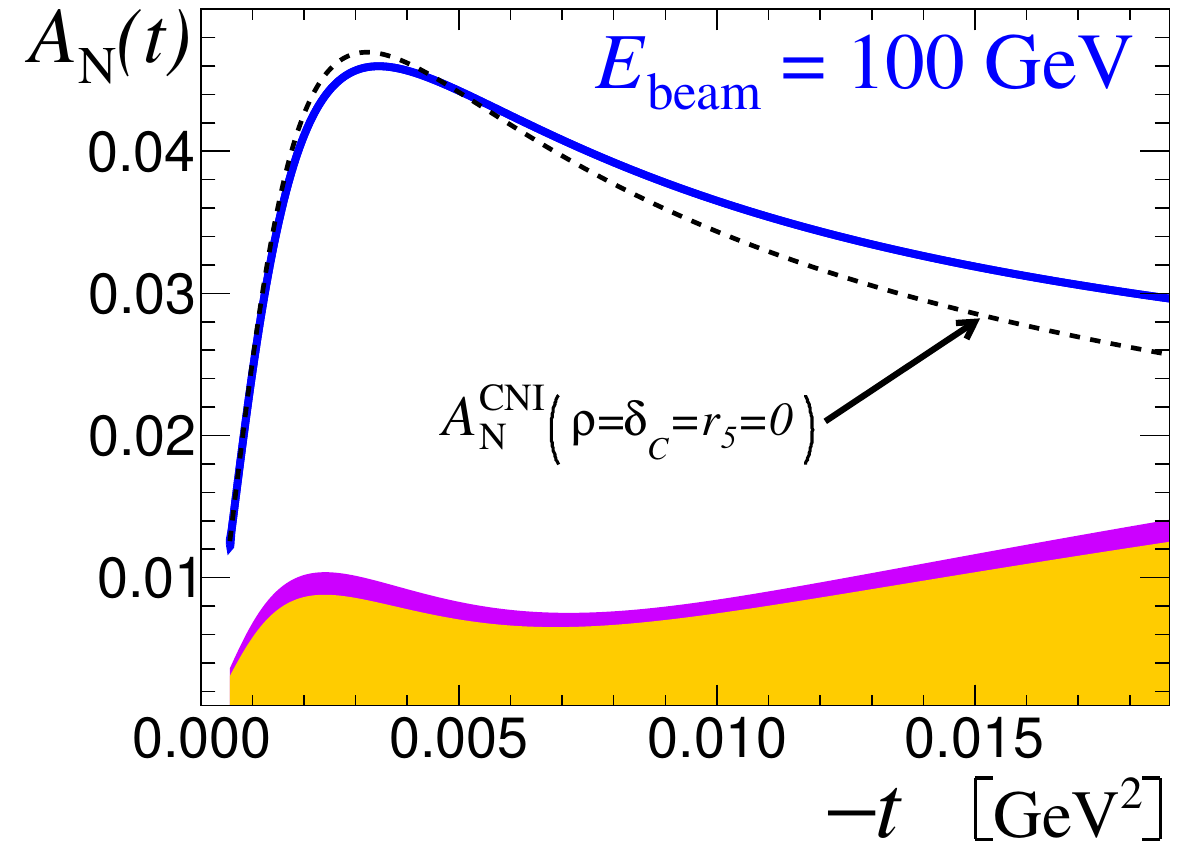}
    \hfill
    \includegraphics[width=0.45\columnwidth]{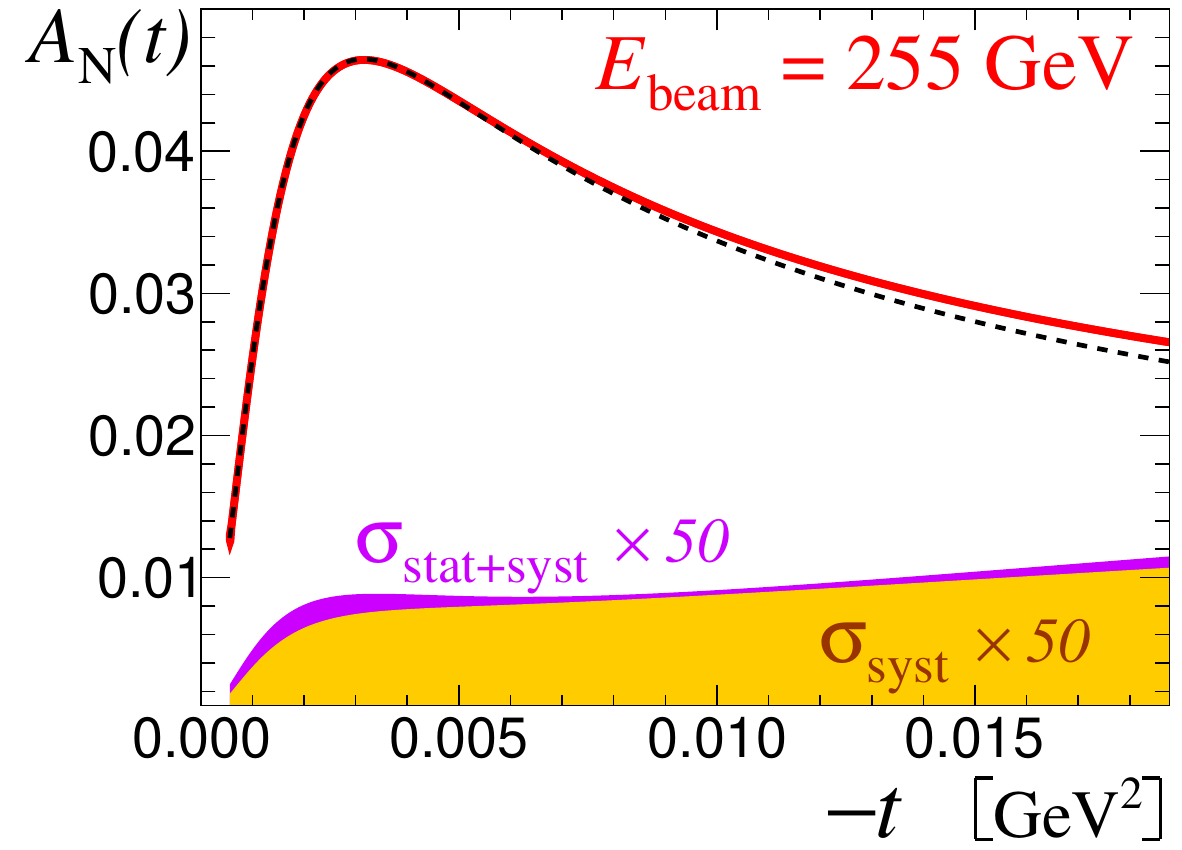}
  \end{center}
  \vspace{-2em}
  \caption{Single-spin analyzing powers for elastic $\mathit{pp}$ scattering measured at HJET \cite{Poblaguev:2019saw}. The filled areas represent 50-fold systematic and total (stat+syst) errors in the measurements. Dashed lines specify the theoretical predictions \cite{Kopeliovich:1974ee} for the Coulomb-nuclear interference only, i.e., neglecting $\rho$, $\delta_C$, and $r_5$ in Eq. (\ref{eq:A_N}).
\label{fig:AN}
}
\end{figure}

The single-spin proton--proton analyzing powers, measured at the HJET\,\cite{Poblaguev:2019saw}, for 100\,GeV ($\sqrt{s}$\,=\,13.76\,GeV) and 255\,GeV ($\sqrt{s}$\,=\,21.92\,GeV) proton beams are displayed in Fig.\,\ref{fig:AN}. Fitting $A_\text{N}(t)$, one finds non-zero hadronic spin-flip amplitudes for both beam energies:    
\begin{align}
  \sqrt{s}=13.76\,\text{GeV}\qquad%
  & R_5=\left(-12.5\pm0.8_\text{stat}\pm1.5_\text{syst}\right)\times10^{-3},
  \label{eq:R5_100}\\
  & I_5=\left(-5.3\pm2.9_\text{stat}\pm4.7_\text{syst}\right)\times10^{-3},
  \\
  \sqrt{s}=21.92\,\text{GeV}\qquad%
  & R_5=\left( -3.9\pm0.5_\text{stat}\pm0.8_\text{syst}\right)\times10^{-3},
  \\
  & I_5=\left( 19.4\pm2.5_\text{stat}\pm2.5_\text{syst}\right)\times10^{-3}.
  \label{eq:I5_255}
\end{align} 
Here, compared to the original publication\,\cite{Poblaguev:2019saw}, the absorption corrections (\ref{eq:R5_abs}) were applied and an updated value of $r_E$\,\cite{Workman:2022ynf} was used. The increase in $|I_5|$ and $|r_5|$ as well as that of the beam energy may be interpreted\cite{Poblaguev:2019saw} as evidence of Pomeron contribution to $r_5$.

\subsection{Energy Dependence of the Hadronic Single Spin-Flip Amplitude}

It is well known that, for unpolarized proton--proton scattering, the elastic amplitude's dependence on the center-of-mass energy squared $s$ can be approximated by
\begin{equation}
  \sigma_\text{tot}(s)\times\left[i+\rho(s)\right] = %
  P(s,\alpha_F) + R^+(s,\alpha_+) + R^-(s,\alpha_-),
\label{eq:Regge_nf}
\end{equation}
where Regge poles
\begin{equation}
  R^\pm(s,\alpha_\pm) \;\propto\;
  \left[1\pm e^{-i\pi\alpha_\pm}\right]\left(s/4m_p^2\right)^{\alpha_\pm-1}
\end{equation}
are encoded as $R^+$ for ($f_2,a_2$) and $R^-$ for ($\omega,\rho$), and a Froissaron parametrization is used for the Pomeron contribution
\begin{equation}
  P(s,\alpha_F) \;\propto\; \pi\alpha_F\ln{(s/4m_p^2)}
  + i\left[1+\alpha_F\ln^2{(s/4m_p^2)}\right].
\end{equation}

In Ref.\,\cite{Poblaguev:2019saw}, to assess the energy dependence of the hadronic spin-flip amplitude, Eq.\,(\ref{eq:Regge_nf}) was reformulated as
\begin{equation}
  \sigma_\text{tot}(s)\times r_5(s) = %
  f_5^P P(s,\alpha_F) + f_5^+ R^+(s,\alpha_+) + f_5^- R^-(s,\alpha_-),
\label{eq:Regge_sf}
\end{equation}
where $f_5^{P,\pm}$ are spin-flip couplings independent of $s$. The parametrization of functions $P(s)$ and $R^\pm(s)$ ($\alpha_F\!=\!0.0090$, $\alpha_+\!=\!0.65$, $\alpha_-\!=\!0.45$) was adopted from Ref.\,\cite{Fagundes:2017iwb}. The Regge pole intercepts $\alpha_\pm$ are assumed to be the same for spin-flip and non-flip amplitudes. However, for the Pomeron, the effective intercepts of its spin-flip and non-flip components may be rather different\,\cite{Kopeliovich:2021rdd,Kopeliovich:1998uy}, potentially invalidating Eq.\,(\ref{eq:Regge_sf}). Nevertheless, due to the limited number of experimental spin-flip entries, this approximation was employed to evaluate the couplings.

The fit of the updated values of $r_5$ (\ref{eq:R5_100})--(\ref{eq:I5_255}) yields
\begin{equation}
  f_5^P = 0.054 \pm 0.002_\text{stat} \pm 0.003_\text{syst},\qquad
  \chi^2/\text{ndf} = 0.7/1.
\end{equation}
This result suggests an unambiguous Pomeron contribution to the spin-flip amplitudes. The obtained value of $\chi^2$ clearly indicates that any improvements to Eq.\,(\ref{eq:Regge_sf}) would be statistically insignificant. However, given that a hypothetical alteration of (\ref{eq:Regge_sf}) could potentially change the value of $f_5^P$, a more conservative conclusion should be drawn: the result of the $r_5(s)$ fit is inconsistent with the contribution from $R^\pm(s)$ alone.

\begin{figure}[t]
  \begin{minipage}[b]{0.43\columnwidth}
    \begin{center}
      \includegraphics[width=1.\columnwidth]{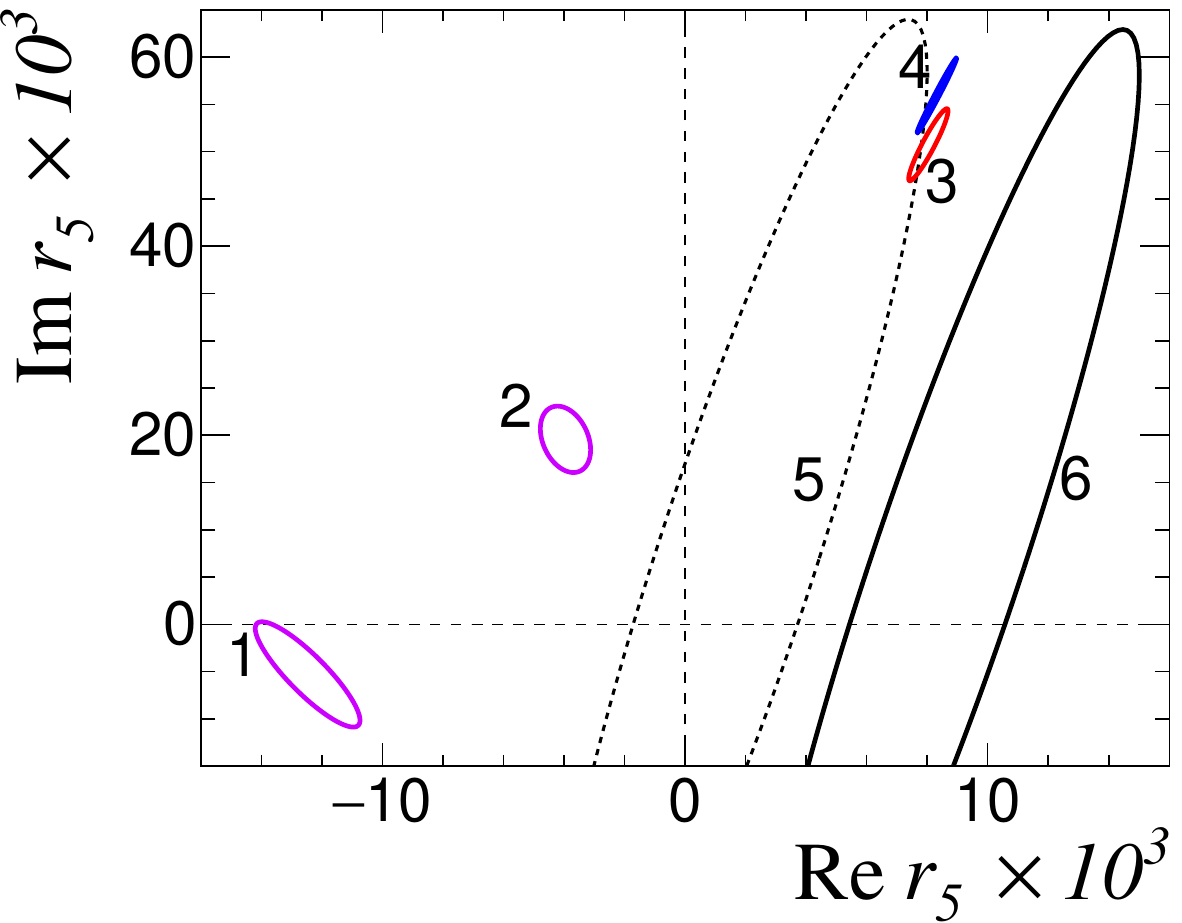}
    \end{center}
  \end{minipage}
  \hfill
  \raisebox{1.2ex} {
    \begin{minipage}[b]{0.53\columnwidth}
      \caption{Experimental 1-sigma contours (stat+syst) for $r_5$. The HJET results are marked as ``1'' ($\sqrt{s}\!=\!13.76$ GeV) and ``2'' (21.92\,GeV). The Regge fit extrapolations to 200\,GeV are labeled as ``3'' for Froissaron and ``4'' for simple pole Pomeron. ``5'' and ``6'' represent the STAR value (200\,GeV) \cite{STAR:2012fiw} before and after applying the absorption and $r_E$-related corrections.
        \label{fig:Regge_r5}
      }
    \end{minipage}
  }
\end{figure}

Using the spin-flip couplings determined in the fit, the value of $r_5$ was evaluated at $\sqrt{s}\!=\!200\,\text{GeV}$ (see Fig.\,\ref{fig:Regge_r5}), enabling a comparison between the HJET and STAR\,\cite{STAR:2012fiw} results. To assess the extrapolation dependence on the model used in the fit, the same analysis was conducted for a simple pole fit with ``standard'' intercepts, $\alpha_{\pm}\!=\!0.5$ and $\alpha_P\!=\!1.10$. The discrepancy between the two extrapolations may be considered insignificant if the resulting value of $r_5$ at $\sqrt{s}\!=\!200\,\text{GeV}$ is compared either with 0 (to test for the Pomeron component) or with the STAR measurement.

Within experimental uncertainties, the HJET measurements align well with the value published\,\cite{STAR:2012fiw} by the STAR Collaboration. However, after applying corrections (\ref{eq:ANcorr})~and~(\ref{eq:R5_abs}), the STAR value was shifted, and consistency between HJET and STAR results may be characterized by $\chi^2/\text{ndf}\!=\!4.8/3$, which is statistically equivalent to 1.8 standard deviations. It must be emphasized that the corrections applied do not include any revision of the analyzing power experimentally determined at STAR.

Although the discrepancy between HJET extrapolation and STAR measurement is not deemed significant, it suggests exploring other possible contributions to the model (\ref{eq:Regge_sf}), e.g., from the Odderon.

For the combined HJET and STAR data, a simple pole Pomeron fit (\ref{eq:Regge_sf}), with the Pomeron intercept $\alpha_P^\text{sf}$ as a free parameter, yields
\begin{equation}
  \alpha_P^\text{sf}=1.13_{-0.03}^{+0.04},\qquad \chi^2/\text{ndf} = 2.8/2,
\end{equation}
which agrees with the unpolarized $\alpha_P^\text{nf}=1.096^{+0.012}_{-0.009}$ \cite{Cudell:1996sh}.

Regardless of HJET measurements being conducted at relatively low energies, \mbox{$\sqrt{s}\!=\!14\text{--}22\,\text{GeV}$,} the result proved to be sensitive to the Pomeron contribution to the proton--proton amplitude. It is also worth noting that the HJET value of $r_5$ extrapolated to $\sqrt{s}\!=\!200\,\text{GeV}$ and the corrected STAR result disagree with recent theoretical evaluations for spin-dependent Pomeron\,\cite{Ewerz:2016onn,Hagiwara:2020mqb,Sawasdipol:2023gro}.

\section{\boldmath Double-Spin Analyzing Power $A_\text{NN}(s,t)$}

\begin{figure}[t]
\begin{center}
\includegraphics[width=0.46\columnwidth]{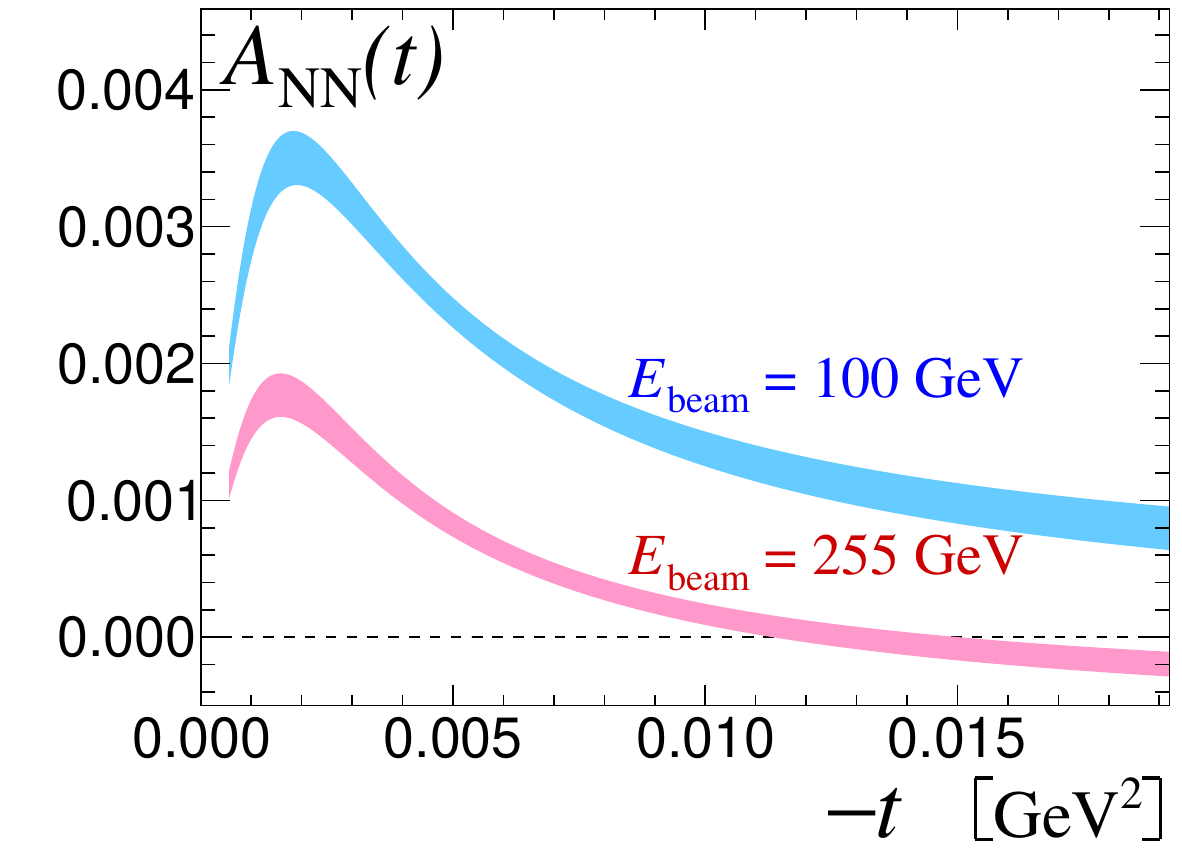}
\hfill
\includegraphics[width=0.46\columnwidth]{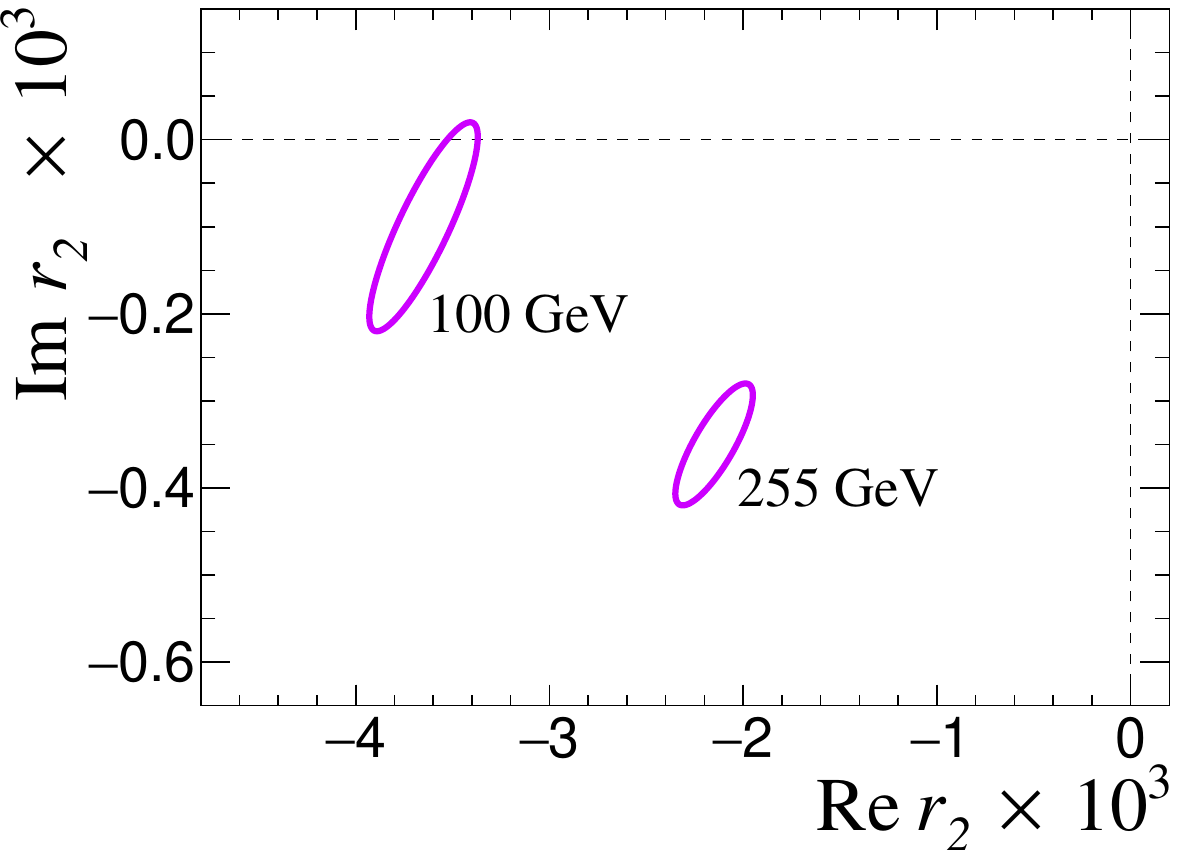}
\end{center}
\vspace{-2em}
\caption{Left: Double spin elastic $p^{\uparrow}p$ analyzing power $A_\text{NN}(t)$ measured at HJET for 100 and 255,GeV proton beams. Right: Double spin-flip amplitude parameters, derived from the measured $A_\text{NN}(t)$. The experimental uncertainties are mainly statistical.
\label{fig:ANN}
}
\end{figure}

For a vertically polarized proton beam and target, the azimuthal distribution of the recoil protons is given\,\cite{Leader_2001} by
\begin{equation}
  \frac{d^2\sigma}{dt\,d\varphi} \propto \left\{
    1+A_\text{N}(t)\sin{\varphi}\left(P_\text{beam}\!+\!P_\text{jet}\right)+
    \left[ A_\text{NN}(t)\sin^2{\varphi}\!+\!A_\text{SS}(t)\cos^2{\varphi} \right]
      P_\text{beam}P_\text{jet}
   \right\}
\end{equation}
where the angle $\varphi\!=\!0$ corresponds to the upward direction. Since for HJET detectors, $\sin{\varphi}\!=\!\pm1$, the measurements are insensitive to the double-spin analyzing power $A_\text{SS}(t)$. Results of the measurement of $A_\text{NN}(t)$ at the HJET are displayed in Fig.\,\ref{fig:ANN}. One can observe that the double spin-flip hadronic amplitude $\phi_2(t)$, parameterized by
\begin{equation}
r_2(s) = \frac{\phi_2^\text{h}(s,t)}{2\mathrm{Im\!\;}\phi_+^\text{h}(s,0)},
\end{equation}
is experimentally well-determined. Using the same approach (\ref{eq:Regge_sf}) as for $r_5(s)$, the Regge fit of $r_2(s)$ gives
\begin{equation}
f_2^P = 0.0020 \pm 0.0002_\text{stat},\qquad
\chi^2/\text{ndf} = 1.6/1,
\end{equation}
i.e., the double spin-flip Pomeron component is statistically well-established.

The sensitivity of $A_\text{NN}(t)$ to the Odderon was discussed in Ref.\,\cite{Leader:1999ua}. The measured $A_\text{NN}(t)$ noticeably disagrees with a theoretical estimate\,\cite{Trueman:2005qd} without Odderon contribution. However, no detailed analysis of the possible Odderon contribution to $A_\text{NN}(t)$, measured at the HJET, has been conducted yet.

\section{Inelastic Proton--Proton Scattering}

\begin{figure}[t]
  \begin{minipage}[c]{0.42\columnwidth}
    \begin{center}
      \includegraphics[width=1.\columnwidth]{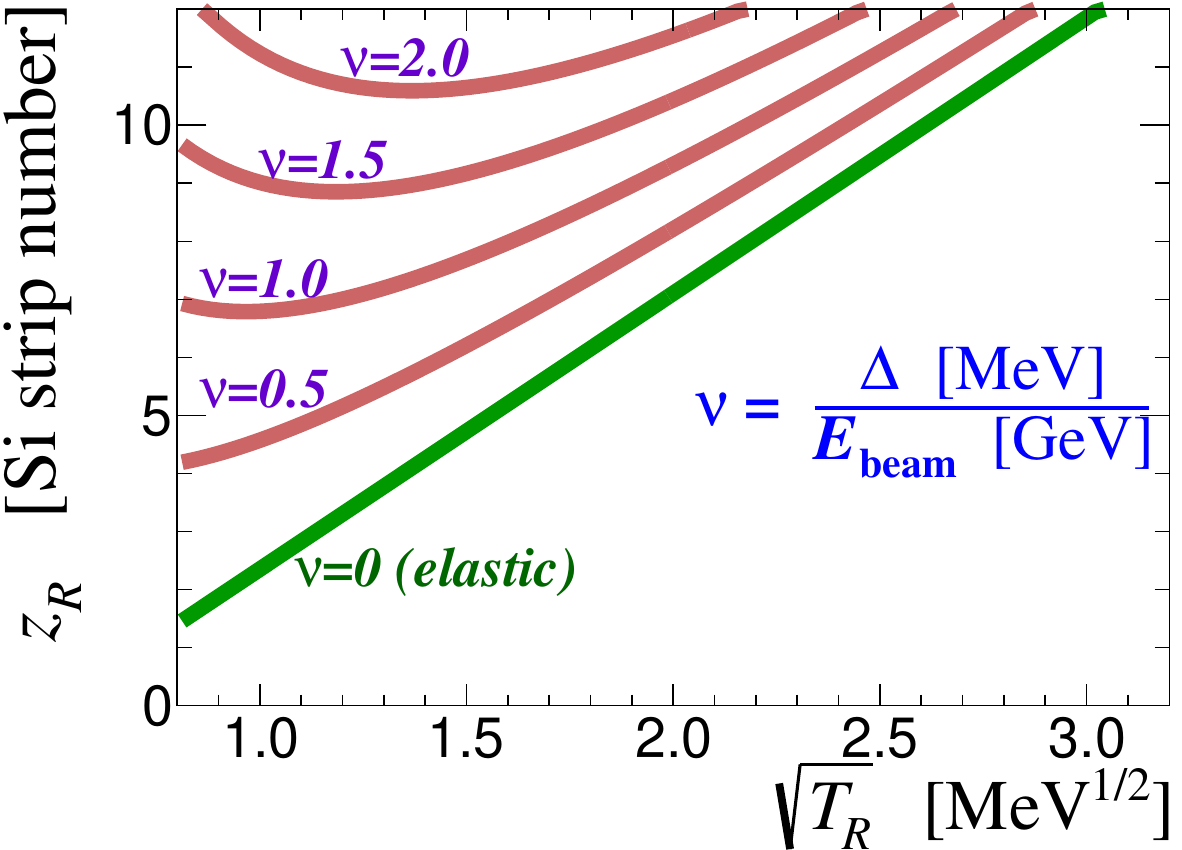}
    \end{center}
  \end{minipage}
  \hfill
  \raisebox{2ex} {
    \begin{minipage}[c]{0.545\columnwidth}
      \caption{
        The recoil proton $z_R(T_R,\Delta)$ coordinate in the Si detector dependence on the kinetic energy $T_R$ and missing mass excess $\Delta$. In terms of $\nu\!=\!\Delta/E_\text{beam}$, the distribution $z_R(T_R,\nu)$ is almost independent of the RHIC beam energy $E_\text{beam}$ per nucleon. Due to the jet thickness, all lines drawn should be smeared with $\sigma_z\!\approx\!0.7$\,strip width. 
        \label{fig:Inelastic}
      }
    \end{minipage}
  }
\end{figure}

Although the scattered beam particle is not detected at the HJET, the measured $z_R$ coordinate of the recoil proton in the Si detector provides some control over the missing mass $M_X$. For inelastic scattering of an ion beam in HJET, the $z_R$ dependence on the missing mass~excess
\begin{equation}
  \Delta = (M_X^2-M^2)/2M \approx M_X - M
\end{equation}
may be approximated by
\begin{equation}
  \frac{z_R-z_\text{jet}}{L} \approx \sqrt{\frac{T_R}{2m_p}}\times\left[
    1 +\frac{m_p^2}{ME_\text{beam}}+\frac{m_p\Delta}{T_RE_\text{beam}} \right],
  \label{eq:zR_inel}
\end{equation}
where $M$ is the ion mass, and $E_\text{beam}$ is the beam energy per nucleon. For elastic ($\Delta\!=\!0$) scattering of a proton ($M\!=\!m_p$) beam, this equation, depicted in Fig.\,\ref{fig:Inelastic}, is consistent with the exact Formula (\ref{eq:zR_el}) with an accuracy of $\sim\!(m_p/E_\text{beam})^2$.

\begin{figure}[t]
  \begin{center}
      \includegraphics[width=0.48\columnwidth]{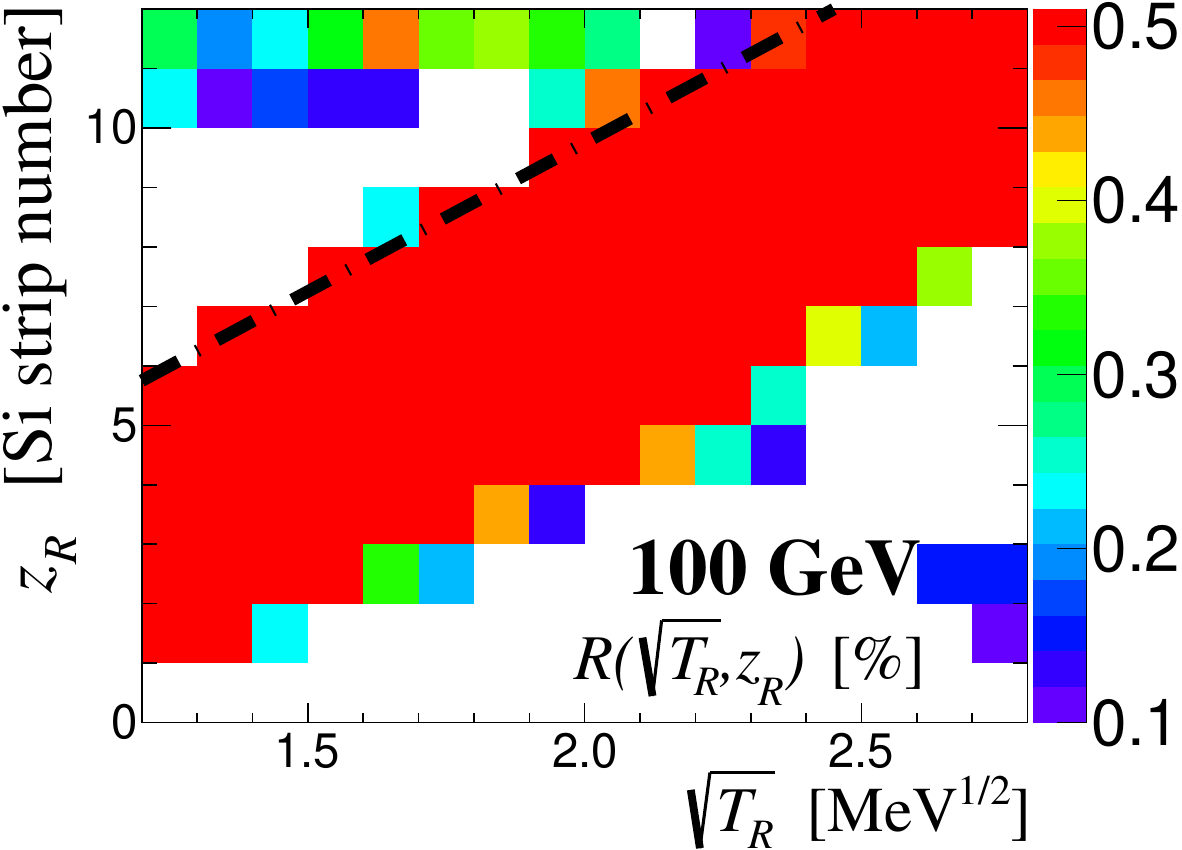}
      \hfill
      \includegraphics[width=0.48\columnwidth]{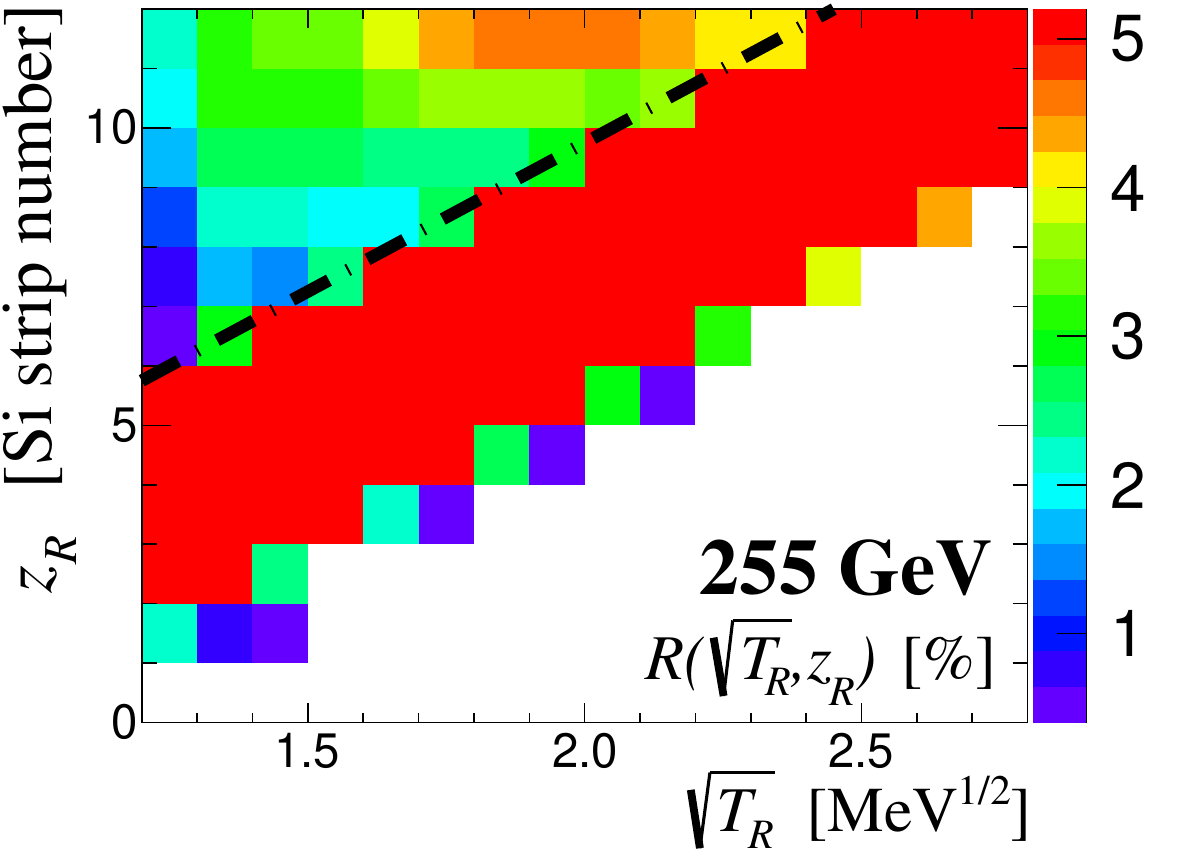}
  \end{center}
  \vspace{-2em}
  \caption{
    Inelastic event rates (above dashed lines) for 100 and 255\,GeV proton beams. The displayed event rate $R(T_R,z_R)$ is normalized by the maximal elastic rate per histogram bin. On the histograms, the rates are cut off at 0.5\% and 5\% for 100 and 255\,GeV, respectively.  
    \label{fig:InelRate}
  }
\end{figure}

\begin{figure}[h]
    \begin{center}
        \includegraphics[width=0.48\columnwidth]{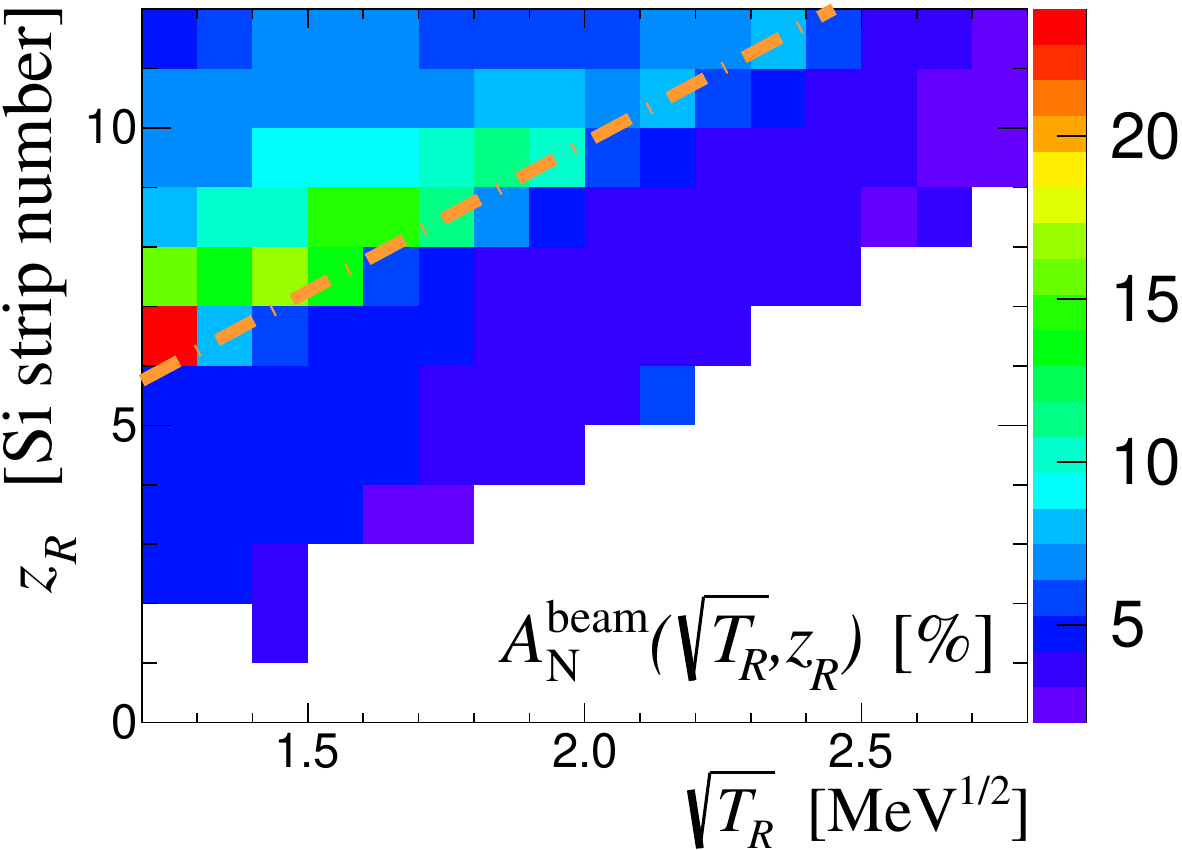}
        \hfill
        \includegraphics[width=0.48\columnwidth]{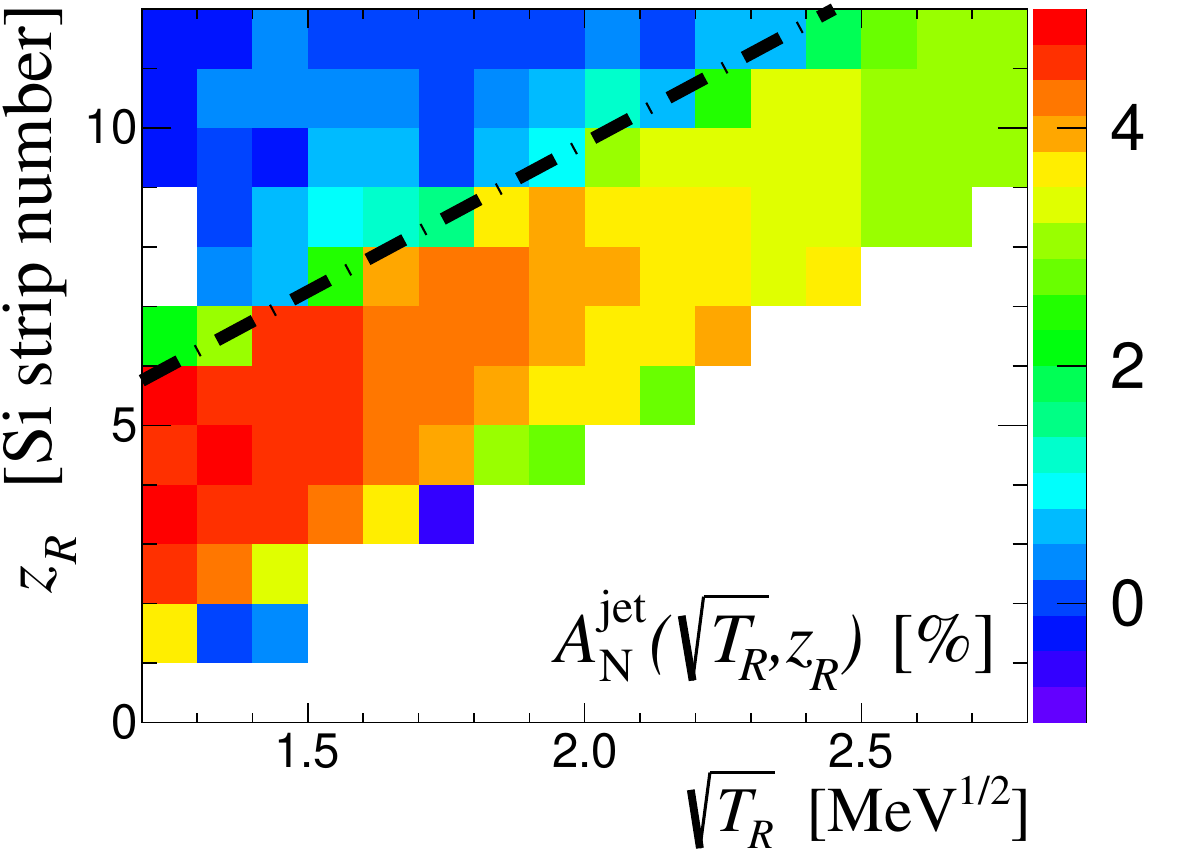}
    \end{center}
    \vspace{-2em}
    \caption{ 
        \label{fig:InelAN_255}
        Inelastic (above dashed lines) beam and jet single spin analyzing powers for the measured 255\,GeV polarized beam scattering of the polarized jet target.   
    }
\end{figure}

\begin{figure}[t]
    \begin{center}
        \includegraphics[width=0.48\columnwidth]{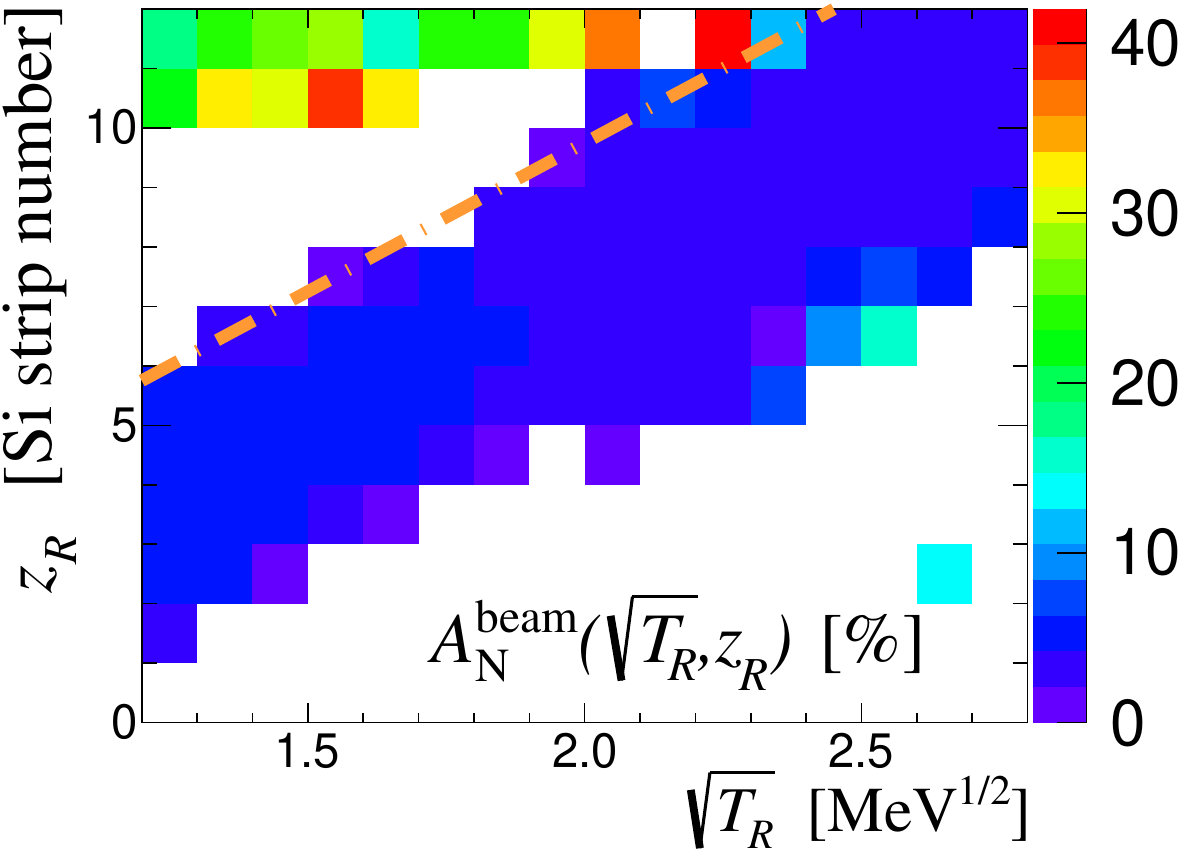}
        \hfill
        \includegraphics[width=0.48\columnwidth]{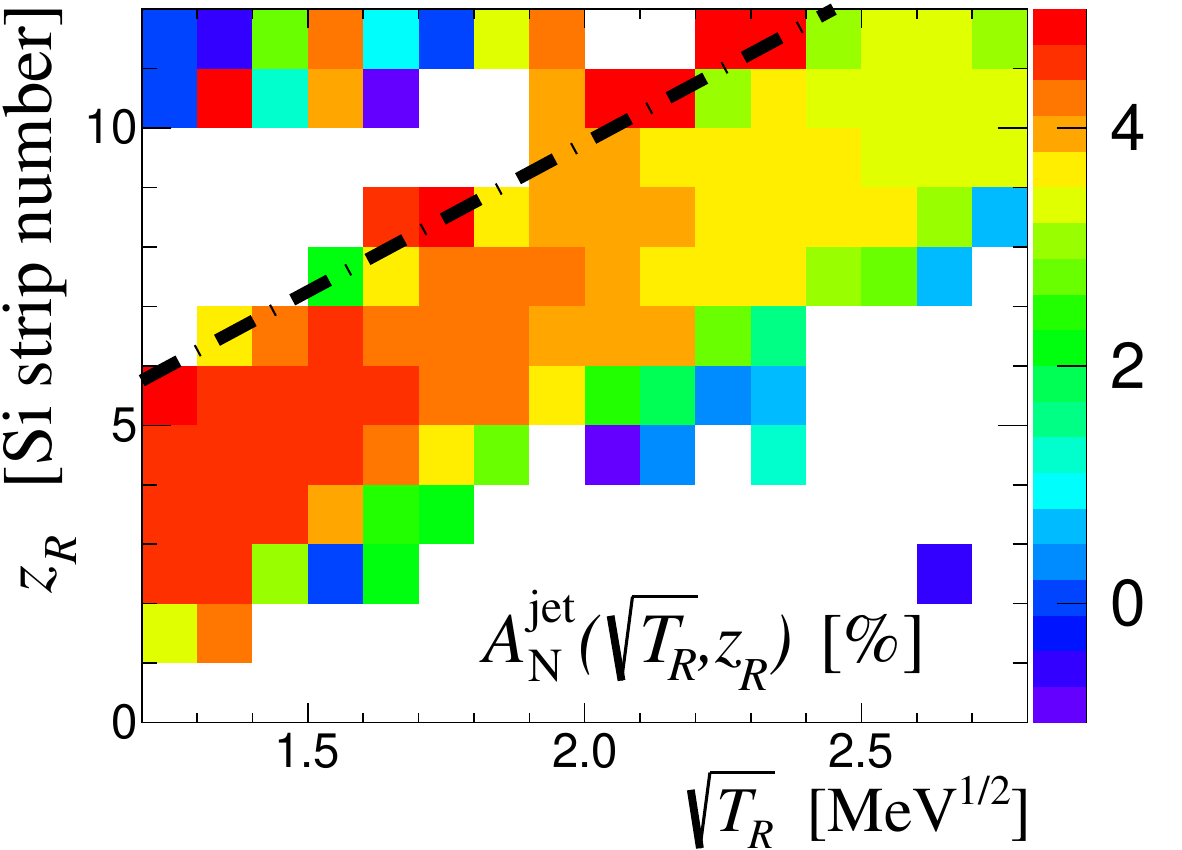}
    \end{center}
    \vspace{-2em}
    \caption{ 
        \label{fig:InelAN_100}
        The same as in Fig.\,\ref{fig:InelAN_255} but for the 100\,GeV proton beam.
    }
\end{figure}

For large values of $\Delta$ ($\nu\!\gtrsim\!2.5$), inelastic events cannot be detected at the HJET. On the other hand, for $\nu\!\lesssim0.9$, inelastic events cannot be separated from the elastic ones.

Fig.\,\ref{fig:InelRate} illustrates the detection of inelastic events for the proton beams. In $pp$ scattering, the inelastic threshold is defined by the pion mass, $\Delta\!>\!m_\pi$. Near this threshold, the event rate is suppressed by the phase space factor, resulting in only a very small fraction of inelastic events being observed in the 100\,GeV data. However, for the 255\,GeV beam, the inelastic rate is significantly larger due to a 2.5 times lower value of $\nu$ corresponding to $m_\pi$.

For inelastic $pp$ events detectable at the HJET, only the beam proton is fragmented. Therefore, the beam $A_\text{N}^\text{beam}(s,t,\Delta)$ and target $A_\text{N}^\text{jet}(s,t,\Delta)$ spin-analyzing powers, which can also depend on $\Delta$, are not necessarily the same in this case.

Preliminary results\,\cite{Poblaguev:2022xoa_} for the beam and target (jet) analyzing powers at 255\,GeV are depicted in Fig.\,\ref{fig:InelAN_255}. Only bins with an event rate $R>0.4\%$ relative to the elastic maximum (i.e., bins displayed in Fig.\,\ref{fig:InelRate}) were considered.
It is evident that, for any value of \(T_R\) at which inelastic analyzing powers are evaluated, 
$A_\text{N}^\text{jet}(t) \lesssim A_\text{N}^\text{elastic}(t) \lesssim A_\text{N}^\text{beam}(t)$.
The inelastic analyzing power increases with decreasing $\Delta$. For the beam spin analyzing power, values of up to 20\% are observed in the data.

Similar conclusions regarding the beam and jet analyzing powers can be drawn for the 100\,GeV beam, as illustrated in Fig.\,\ref{fig:InelAN_100}. Since events with lower values of $\Delta$ can be separated from the elastic data in comparison to the 255\,GeV measurements, the observed beam spin analyzing power is somewhat larger, reaching up to 30\%. However, the statistical significance is considerably smaller, and the measurements are less accurate.

\section{Elastic Proton--Nucleus Analyzing Power}

\begin{figure}[t]
  \begin{center}
    \includegraphics[width=0.48\columnwidth]{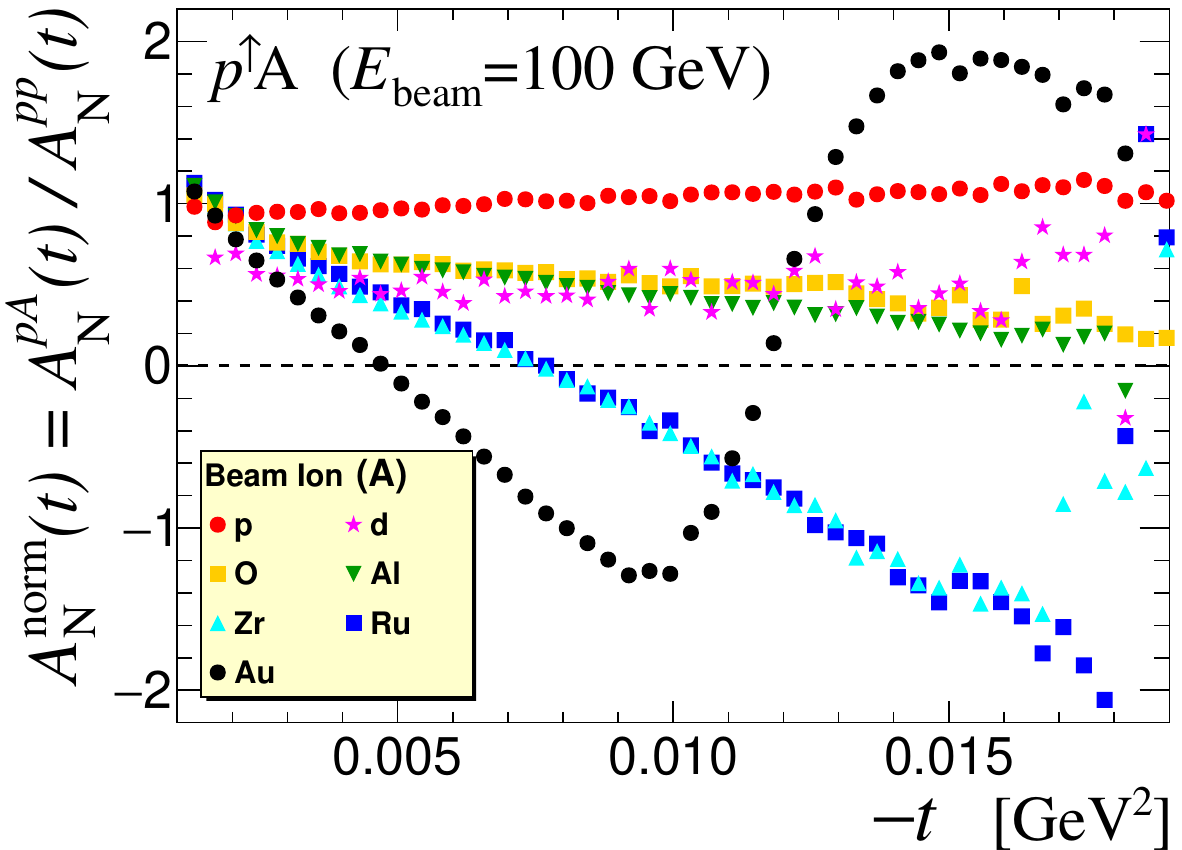}
    \includegraphics[width=0.48\columnwidth]{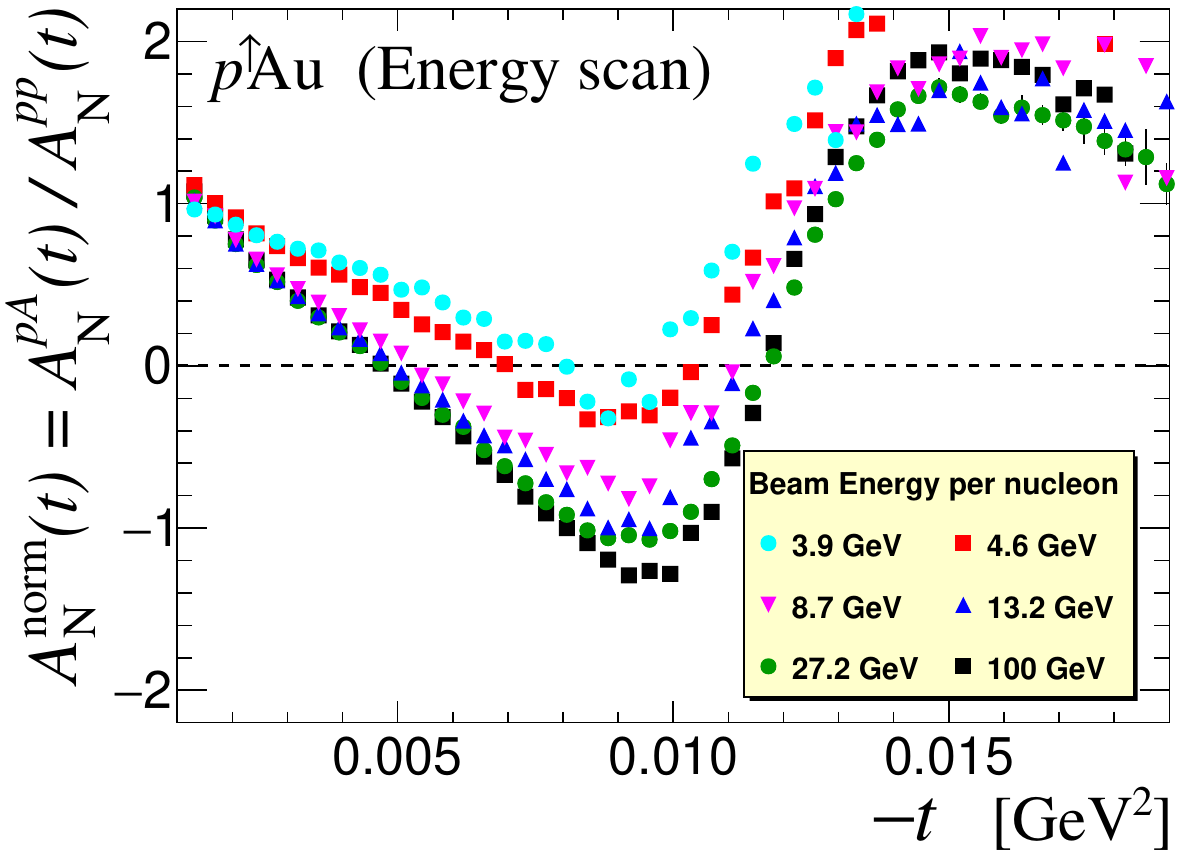}
  \end{center}
  \vspace{-2em}
  \caption{
    Dependence of the proton-nucleus elastic analyzing power $A_\text{N}^{pA}(t)$ on the beam ion (left) and the beam energy (right). The measured analyzing powers are normalized by the proton-proton one calculated for $E_\text{beam}\!=\!100\,\text{GeV}$, assuming no hadronic single spin-flip ($r_5=0$).
    \label{fig:pA}
  }
\end{figure}

Since 2015, the HJET has been routinely operated during RHIC Heavy Ion Runs, demonstrating consistent performance with both ion and proton beams. This operational stability enables the precise measurements of the proton--nucleus analyzing power $A_\text{N}^{pA}(t)$. A study of the analyzing powers has been conducted for six ions ($^2$H$^{+}$, $^{16}$O$^{8+}$, $^{27}$Al$^{12+}$, $^{96}$Zr$^{40+}$, $^{96}$Ru$^{44+}$, and $^{197}$Au$^{79+}$), including beam energy scans for Au and deuteron ($d$) beams. Preliminary results\,\cite{Poblaguev:2022xoa_} are presented in Fig.\,\ref{fig:pA}. Notably, systematic uncertainties in the measurements were not assessed in this analysis.

For the 100\,GeV/nucleon Au beam, experimental data were compared with theoretical predictions in Ref.\,\cite{Krelina:2019mlu}. The study emphasized the importance of absorption corrections in calculating $A_\text{N}^{p\text{Au}}(t)$. However, not all notable discrepancies between the data and theory were fully resolved.
A new theoretical analysis of the $p^{\uparrow}A$ analyzing power at 100\,GeV is presented in Ref.\,\cite{Kopeliovich:2023xtu}.

From a kinematic perspective, ion beam breakup events can be effectively isolated in the conducted studies due to the low threshold for the missing mass excess, $\Delta \lesssim 10\,\text{MeV}$. However, across numerous measurements spanning a wide range of beam energies, no evidence of breakup events was found in the data, except for a small admixture, approximately 1\%, observed in the elastic data for deuteron beams at lower energies, $\leq\!\!30\,\text{GeV/nucleon}$.

To experimentally assess the fraction of breakup events in the gold beam measurements, a dedicated study was conducted using single RHIC beams at energies of 3.85 and 26.5\,GeV/nucleon, with the HJET holding field magnet turned off. This configuration significantly reduced background-related uncertainties. For the recoil proton energy range $1.3\!<\!\sqrt{T_R}\!<\!2.1\,\text{MeV}^{1/2}$ ($0.003\!<\!-t\!<\!0.009\,\text{GeV}^2$), the following preliminary estimates of the breakup fraction $\langle\sigma_\text{brk}^{p\text{Au}}/\sigma_\text{el}^{p\text{Au}}\rangle$ in the detected events, averaged over $T_R$ and $\Delta$, were obtained\,\cite{Au_Breakup}:
\begin{align}
  &\text{3.85 GeV:} \quad \quad0.20 \pm 0.12\,\% \quad [3.6\!<\!\Delta\!<\!8.5\,\text{MeV}], \\
  &\text{26.5 GeV:} \quad -\!0.08 \pm 0.06\,\% \quad [20\!<\!\Delta\!<\!60\,\text{MeV}]. 
\end{align}
An explanation of the obtained result will be provided below.

Also, taking into account that, for a breakup $p^{\uparrow}A$ amplitude, the spin-flip parameter $r_5$ is expected to be the same as for the elastic amplitude\,\cite{Poblaguev:2023gfl}, one should conclude that the $p^{\uparrow}A$ analyzing power measured at the HJET should be mainly interpreted as elastic.

\section{\boldmath $^3$He Beam Polarization Measurement}

Polarized $^3$He ($Z_h\!=\!2$, $A_h\!=\!3$) beams are planned at the EIC. The possibility of measuring the beam polarization with the required accuracy (\ref{eq:EIC_syst}) was investigated in Refs.\,\cite{Poblaguev:2022gqy,Poblaguev:2022hsi,Poblaguev:2023gfl}. Since HJET operation in ion beams is already well understood, it was suggested to measure the helion beam polarization in the same way as it was done (\ref{eq:Pbeam}) for the proton beam. In this case, however, analyzing powers for $p^{\uparrow}h$ and $h^{\uparrow}p$ scattering are not the same. Consequently, the $A_\text{N}^{ph}(t)/A_\text{N}^{hp}(t)$ ratio factor should be added to the right-hand side of Eq.\,(\ref{eq:Pbeam}), leading to
\begin{equation}
  P_\text{meas}^h(T_R) = P_\text{jet}\frac{a_\text{beam}(T_R)}{a_\text{jet}(T_R)}%
  \times \frac%
         {\kappa_p-2I_5^{ph}-2R_5^{ph}T_R/T_c}{\kappa_h-2I_5^{hp}-2R_5^{hp}T_R/T_c},
         \label{eq:Pmeas}
\end{equation}
where $\kappa_h\!=\mu_h/Z_h\!-\!m_p/m_h\!=\!-1.398$, $\mu_h$ is the magnetic moment of $^3$He, and $T_c\!=\!4\pi\alpha{Z_h}/m_p\sigma_\text{tot}^{ph}\!\approx\!0.7\,\text{MeV}$. For simplicity, only the dominant components of the interference terms (\ref{eq:AN_phi}) are shown in Eq.\,(\ref{eq:Pmeas}).

Anticipating errors in values of $r_5^{ph}$ and $r_5^{hp}$, one can expect dependence of the measured polarization on the recoil proton energy
\begin{equation}
  P_\text{meas}^h(T_R) = P_\text{beam}^h\times\left[ 1 +\xi_0+\xi_1T_R/T_c+\dots\right],
  \label{eq:t->0}
\end{equation}
where $P_\text{beam}^h$ is the actual $^3$He beam polarization, and $\xi_{0,1}$ parameterize systematic uncertainties in the polarization measurement. Thus, the extrapolation of $T_\text{beam}^h(T_R)$ to $T_R\!\to\!0$ allows one to eliminate the effect of the systematic uncertainties in real parts (which are enhanced by a factor of $T_R/T_c$) of the hadronic spin-flip amplitudes.

\subsection{Analyzing Power for $p^{\uparrow}h$ and $h^{\uparrow}p$ Scattering}

In Ref.\,\cite{Kopeliovich:2000kz}, it was shown that, at high energy, to a very good approximation, the ratio of the spin-flip to the non-flip parts of the elastic proton--nucleus amplitude is the same as for proton--nucleon scattering, i.e., 
\begin{equation}
  r_5^{pA} = \frac{i+\rho^{pA}}{i+\rho^{pp}}r_5\approx r_5,
  \label{eq:r5_pA}
\end{equation}
where $\rho^{pA}$ and $\rho^{pp}$ are the real-to-imaginary ratios for elastic $pA$ and $\mathit{pp}$ scattering, respectively. This result can be readily derived considering the polarized proton scattering of an unpolarized nucleus in Glauber (diffraction) approximation\,\cite{Glauber:1955qq,Glauber:1959}.

For proton--deuteron small-angle scattering, the elastic $pd$ amplitude $F_{ii}$ can be approximated\,\cite{Franco:1965wi} by combining the proton--proton $f_p$ and proton--neutron $f_n$ ones as 
\begin{equation}
  F_{ii}(\bm{q}) =
  S(\bm{q}/2)f_n(\bm{q}) +  
  S(\bm{q}/2)f_p(\bm{q}) +
  \frac{i}{2\pi k}\int{S(\bm{q}')f_n(\bm{q}/2\!+\!\bm{q}')f_p(\bm{q}/2\!-\!\bm{q}')d^2\bf{q}'},
  \label{eq:Fii_pd}
\end{equation}
where $S(\bm{q})$ can be interpreted as a deuteron form factor.

To calculate the $p^\uparrow d$ spin-flip amplitude $F_{ii}^\text{sf}$, one can utilize the proton--nucleon one, $f_N^\text{sf}(\bm{q})$,  which, according to the definition of $r_5$ (\ref{eq:r5_def}), is
\begin{equation}
  f_N^\text{sf}(\bm{q})=\frac{\bm{qn}}{m_p}\,\frac{r_5^{pp}}{i+\rho^{pp}}\,f_N(\bm{q})
  = \bm{qn}\,\hat{r}_5^{pp}f(\bm{q}).
\end{equation}
Since $|r_5^{pp}|\,q/m_p\lesssim0.003$ in the HJET measurements, only one non-flip amplitude $f_N$ in each term of Eq.\,\eqref{eq:Fii_pd} should be replaced by its spin-flip counterpart $f_N^\text{sf}$. In particular,
\begin{equation}
  f_pf_n \to \left[(\bm{q}/2\!+\!\bm{q}')\bm{n}f_p\,f_n + f_p\,(\bm{q}/2\!-\!\bm{q}')\bm{n}f_n\right]\,\hat{r}_5^{pp} =
 \bm{qn}\,\hat{r}_5^{pp}\,f_pf_n.
\end{equation}
Due to the fact that each term on the right side of Eq.\,(\ref{eq:Fii_pd}) acquires a factor $\bm{qn}\,\hat{r}_5$, one can readily arrive at Eq.\,(\ref{eq:r5_pA}) for the deuteron. The proof is easily extendable for an arbitrary nucleus\,\cite{Poblaguev:2023gfl}, and for elastic $p^{\uparrow}h$ scattering, one finds, with good accuracy, $r_5^{ph}\!=\!r_5^{pp}$.

In the case of unpolarized proton elastic scattering from a fully polarized spin-1/2 nucleus with a space-symmetric distribution of nucleons (e.g., from $^3$He), the proton-nucleus spin-flip parameter is evidently proportional to the average polarization of the nucleons, $r_5^{Ap} \cong r_5\sum{P_i}/A$. Assuming that $^3$He nuclei are in a space-symmetric ${}^1S_0$ state, with the helion spin carried by the neutron (i.e., $P_n = 1$ and $P_p = 0$), one obtains $r_5^{hp} = r_5^{pp}/A_h$ \cite{Buttimore:2001df}.

In a more detailed analysis \cite{Poblaguev:2022gqy}, the following estimates were obtained:
\begin{equation}
  r_5^{ph} = r_5^{pp} \pm 0.001 \pm i\,0.001,\qquad%
  r_5^{hp} = 0.27\!\;r_5^{pp} \pm 0.001 \pm i\,0.001.
\end{equation}
These results give rise to systematic uncertainties in the measured $^3$He beam polarization as follows:
\begin{equation}
 \!\!\!P_\text{beam}^h = P_\text{meas}^h(0)\times%
  \left[1\pm0.006_\text{syst}\pm0.005_{r_5}\pm0.002_\text{mod}\right]\!.
  \label{eq:PhErrs}
\end{equation}
Here, the subscript ``syst'' denotes systematic uncertainty in the measured $a_\text{beam}/a_\text{jet}$ asymmetry ratio, ``$r_5$'' accounts for experimental error in the pre-determined proton-proton $r_5^{pp}$, and ``mod'' specifies the model-dependent theoretical accuracy of the relations between proton--proton and helion--proton values of $r_5$.

\subsection{Breakup Corrections for the Measured $^3$He Beam Polarization}

Potentially, results of the $^3$He beam polarization measurement may also be affected by the helion breakup in the scattering. Since the $^3$He binding energy is only 5.5\,MeV and the EIC beam energy will be of the order of 100\,GeV, a naive expectation is that the elastic data in the polarization measurements will be strongly contaminated by the breakup events. 

However, as mentioned earlier, no evidence of breakup events was found in numerous HJET measurements with ion beams, except for deuteron beams at the 10--30\,GeV/nucleon. In this specific case, the breakup fraction, which can be experimentally isolated at the HJET, is only about 1\% compared to the elastic event statistics.

This phenomenon can be explained using a simple model\,\cite{Poblaguev:2022hsi}. If the breakup, in a high-energy proton beam scattering of a nucleus target, is caused by incoherent scattering of a nucleon in the nucleus or, more generally, off a nucleon cluster with mass $m^*$, a simple kinematical consideration gives for the missing mass excess $\Delta$, 
\begin{equation}
  \Delta = \left(1-\frac{m^*}{M_A}\right) T_R + p_x\sqrt{\frac{2T_R}{m_p}}
  \label{eq:Delta}
\end{equation}
where $p_x$ is the transverse momentum of the cluster due to internal motion of nucleons in the nucleus. If $p_x$ has a Breit--Wigner distribution
\begin{equation}
  dN/dp_x \propto f_\text{BW}(p_x,\sigma_p) = %
  \frac{\pi^{-1}\sqrt{2}\sigma_p}{p_x^2\!+\!2\sigma_p^2},
\end{equation}
then the breakup event rate dependence on $\Delta$ may be approximated by
\begin{equation}
  dN/d\Delta \propto f_\text{BW}(\Delta-\Delta_0,\sigma_\Delta)\Phi_2(\Delta),
\end{equation}
where $\Delta_0\!=\!(1\!-\!m^*/M_A)T_R$, $\sigma_\Delta\!=\!\sigma_p\sqrt{2T_R/m_p}$, and $\Phi_2(\Delta)$ is a phase space factor (here, for the 2-body breakup, since other modes are more strongly suppressed for small $\Delta\!\ll\!M_A$).

The model was successfully tested\,\cite{Poblaguev:2022hsi,Poblaguev:2023gfl} using HJET data obtained with 10--30\,GeV (per nucleon) deuteron beams, and a value of $\sigma_p\!\approx\!30\,\text{MeV}$ was found for the deuteron.

\begin{figure}[t]
  \begin{minipage}[c]{0.55\columnwidth}
    \begin{center}
      \includegraphics[width=1.0\columnwidth]{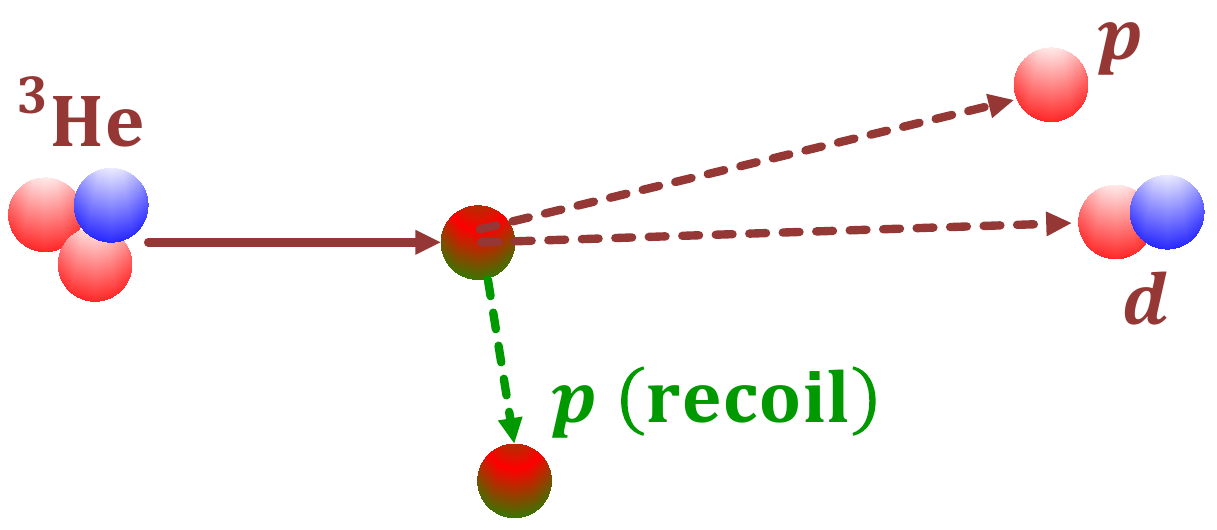}
    \end{center}
  \end{minipage}
  \hfill
  \begin{minipage}[c]{0.42\columnwidth}
    \caption{Dominant breakup mode $h\!\to\!pd$ for the EIC $^3$He beam scattering, detectable at the HJET.
      \label{fig:Breakup}
    }
  \end{minipage}
\end{figure}

Since only low-energy events with $T_R\!<\!10\,\text{MeV}$ ($-t\!<\!0.02\,\text{GeV}^2$) occur in the HJET, the missing mass excess for the breakup events is constrained by $\Delta\!\lesssim\!10\,\text{MeV}$. Consequently, (i) the detection rate of breakup events is significantly suppressed by the space phase factor, (ii) for the $^3$He breakups observable in the HJET, the 2-body mode $h\!\to\!pd$ (see Fig.\,\ref{fig:Breakup}) dominates prominently, and (iii) with the 100\,GeV/nucleon helion beam, practical separation of breakup events from elastic ones is not achievable\,\cite{Poblaguev:2023qln}.

\begin{figure}
  \begin{minipage}[c]{0.55\columnwidth}
    \begin{center}
      \includegraphics[width=1.0\columnwidth]{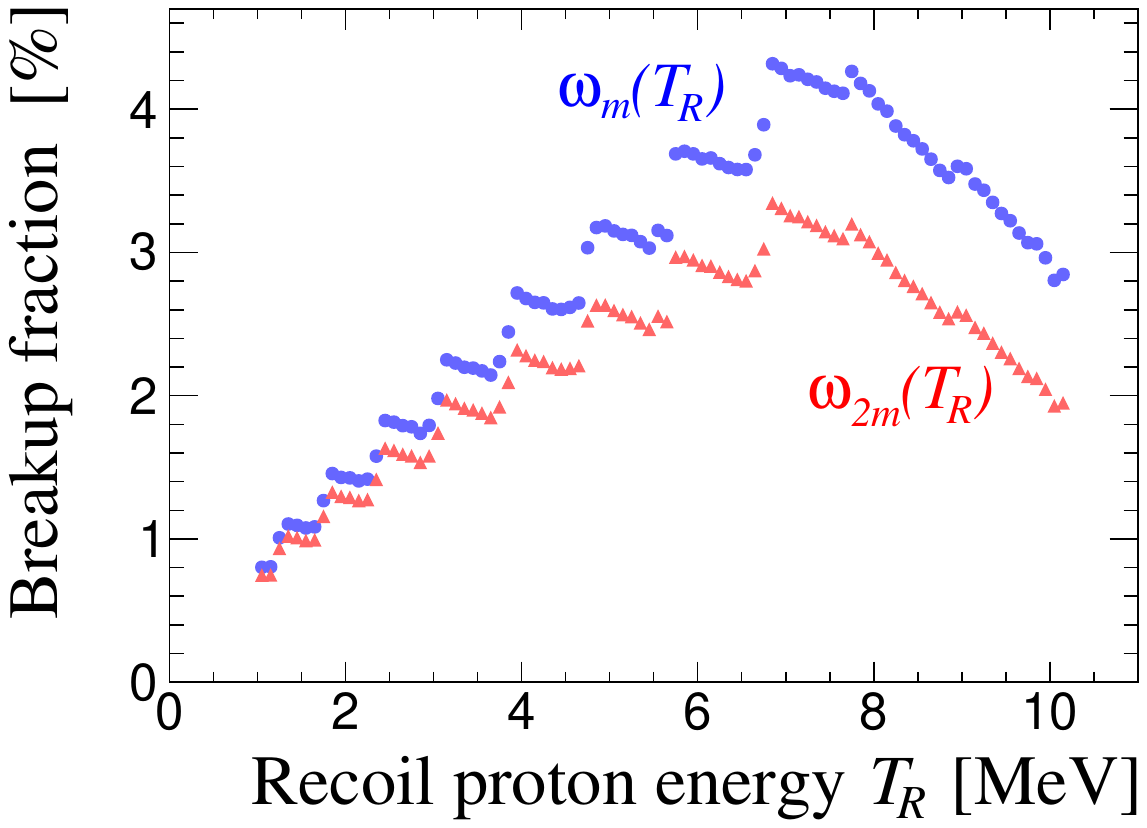}
    \end{center}
  \end{minipage}
  \hfill
  \raisebox{3ex} {
  \begin{minipage}[c]{0.42\columnwidth}
    \caption{
      Anticipated $^3$He beam breakup fraction in the HJET measurements at EIC. The estimates, based on the experimental study of the deuteron beam breakup, have been carried out for $\sigma_p\!=\!30\,\text{MeV}$ and two values of the nucleon cluster mass [Eq.\,(\ref{eq:Delta})], $m^*\!=\!m_p$ and $m^*\!=\!2m_p$.
      \label{fig:omega}
    }
  \end{minipage}
  }
\end{figure}

If the detection of the scattered $^3$He beam is unavailable, the breakup can be identified by an effective alteration of the elastic cross-section:
\begin{equation}
  (d\sigma/dt)_\text{el} \to (d\sigma/dt)_\text{el} \times \left[ 1+\omega(t) \right]
\end{equation}
where $\omega(t)$ represents the breakup fraction in the elastic data. Since $\omega(0)\!=\!0$ (unless the nucleus can spontaneously decay), one can expect that $\omega(t)$ linearly depends on $t$ (or $T_R$) for small momentum transfer squared. Based on the deuteron beam breakup study at HJET\,\cite{Poblaguev:2022hsi,Poblaguev:2023gfl}, the function $\omega(T_R)$ was evaluated (see Fig.\,\ref{fig:omega}) for the 100\,GeV $^3$He beam, taking into account the elastic event selection cuts and the discreteness of the $z_R$ coordinate determined by the Si strip number. The obtained $\omega_m(T_R)$ should be interpreted\,\cite{Poblaguev:2023gfl} as an upper limit for the $^3$He breakup fraction in the HJET measurements.

In Eq.\,(\ref{eq:Pmeas}), analogous breakup corrections $\omega$, generically denoted as $\omega_\text{int}(T_R)$, should be applied to each interference term
\begin{equation}
  \frac{\kappa_p - 2I_5^{ph} - 2R_5^{ph}T_R/T_c}%
       {\kappa_h - 2I_5^{hp} - 2R_5^{hp}T_R/T_c}%
  \quad\to\quad \frac%
      {\kappa_p[1+\omega_\kappa^p] - 2I_5^{ph}[1+\omega_I^p] -
        2R_5^{ph}[1+\omega_R^p]\,T_R/T_c}%
      {\kappa_h[1+\omega_\kappa^h] - 2I_5^{hp}[1+\omega_I^h] -
        2R_5^{hp}[1+\omega_R^h]\,T_R/T_c}.
\end{equation}
Since these corrections vanish as $T_R\to 0$, in accordance with Eq.\,(\ref{eq:t->0}), they should not impact the measured polarization. However, if the hadron beam polarization can be measured only for $T_R > 2\,\text{MeV}$ \cite{Poblaguev:2020Og}, an analysis of the extrapolation of the measured values of $P_\text{meas}^h(T_R)$ to $T_R\!=\!0$ becomes necessary.

Substituting $t$ and $M_X^2$ by $T_R$ and $\Delta$, a breakup amplitude can be written as
\begin{equation}
  \phi_\text{brk}(T_R,\Delta) = \phi_\text{el}(T_R)\phi_\text{BW}(T_R,\Delta)\widetilde{\phi}(T_R,\Delta) = \phi_\text{el}(T_R)\phi_\text{BW}(T_R,\Delta)\widetilde{\phi}_0,
\end{equation}
where $\phi_\text{el}(T_R)$ is the corresponding elastic amplitude, and
\begin{equation}
  \left|\phi_\text{BW}(T_R,\Delta)\right|^2=f_\text{BW}(\Delta-\Delta_0,\sigma_\Delta).
\end{equation}
Since the momentum transfer squared dependence of $\phi_\text{brk}$ is defined by $\phi_\text{el}(T_R)$ and the dependence on $\Delta$ is given by $\phi_\text{BW}(T_R,\Delta)$, for the HJET measurements, one can expect only a weak dependence of $\widetilde{\phi}$ on $T_R$ and $\Delta$. Therefore, $\widetilde{\phi}(T_R,\Delta)$ can be replaced by a constant $\widetilde{\phi}_0\!=\!\widetilde{\phi}(0,0)$.

For the $h\!\to\!pd$ breakup, a value of $\widetilde{\phi}_0$ is defined by helion and proton--deuteron wave functions in rest. Therefore, it should be the same for all considered amplitudes, spin-flip/non-flip, and hadronic/electromagnetic. For example, since $f_\text{BW}(T_R,\Delta)$ is the same for spin-flip and non-flip hadronic $p^{\uparrow}h$ amplitudes, the inelastic spin-flip parameter $r_5^\text{brk}$, if it is defined
\begin{equation}
  \phi_5^\text{brk} = \frac{\sqrt{-t}}{m_p}\frac{r_5^\text{brk}}{i+\rho^\text{el}}\phi_+^\text{brk}
\end{equation}
using the elastic value of the real-to-imaginary ratio $\rho^\text{el}$, should be equal to the elastic $r_5^{ph}$, and, consequently, $\omega_R^p(T_R)\!=\!\omega(T_R)$. Such a conclusion can also be derived directly from Eq.\,(\ref{eq:Fii_pd}). Since this equation is also valid for the breakup amplitude $F_{if}$ (but with some different form factor(s) $S({\bf q})$), the same analysis as that carried out for the elastic amplitude also leads to $r_5^{ph\text{(brk)}}=r_5^{ph\text{(el)}}$.

In a more general case, $f_\text{BW}(T_R,\Delta)$ may depend on the amplitude type. For instance, the electromagnetic spin-flip amplitude for $p^{\uparrow}h$ scattering involves the beam proton interacting with two protons in $^3$He. In contrast, for unpolarized proton scattering of a polarized helion, the amplitude is defined by interaction with the polarized neutron in $^3$He. Nevertheless, given that the sum of $p_x$ for all nucleons in the nucleus must be zero, one can expect that, for $^3$He, with a symmetric wave function, the $p_x$ distribution for a di-nucleon is the same as for any nucleon. If so, all $\omega_\text{int}(T_R)$ are expected to be constrained by the breakup fraction functions calculated for $m^*\!=\!m_p$ and $m^*\!=\!2m_p$:
\begin{equation}
  \omega_{2m}(T_R)\le\omega_\text{int}(T_R)\le\omega_m(T_R).
\end{equation}
Due to there being only a small difference between $\omega_m(T_R)$ and $\omega_{2m}(T_R)$ (see Fig.\,\ref{fig:omega}), the breakup corrections should essentially cancel in the $^3$He beam polarization measurement at the HJET.

For a test, the impact of potential discrepancies between $\omega_\text{int}(T_R)$ values can be emphasized by setting $\omega_\text{int}\!=\!\omega_m$ for $p^{\uparrow}h$ scattering and $\omega_\text{int}\!=\!\omega_{2m}$ for $h^{\uparrow}p$. In this scenario, a linear fit of the breakup corrections in the recoil proton energy range of $2\!<\!T_R\!<\!10\,\text{MeV}$ yields the following errors in
\begin{equation}
  P_\text{meas}(T_R) = P_\text{beam}^h\times\left[1 -0.11\% + 0.13\%\,\frac{T_R}{T_c}\right].
\end{equation}
The relatively large value of the systematic error at $T_R\!=\!0$ (irrespective of whether it satisfies requirement (\ref{eq:EIC_syst})) can be attributed to the strong non-linearity of $\omega_\text{int}(T_R)\,T_R/T_c$. In a more accurate data processing\,\cite{Poblaguev:2023gfl}, this systematic error can be mitigated.

Therefore, although the breakup correction to the interference terms in Eq.\,(\ref{eq:Pbeam}) may be as large as 4\% (relative), the effect strongly cancels in the $p^{\uparrow}h$ and $h^{\uparrow}p$ analyzing power ratio. Consequently, it is expected to be negligible in the EIC $^3$He beam polarization measurement with HJET.

\section{Summary}

The Polarized Atomic Hydrogen Gas Jet Target (HJET) polarimeter was initially designed to measure the absolute proton beam polarization at RHIC with an accuracy better than 5\%. Over two decades of operation, the HJET demonstrated superior performance, achieving systematic errors in proton beam polarization measurements of $\sigma_P/P \lesssim 0.5\%$.

The well-controlled background and low systematic uncertainties in the measurements facilitated a precision experimental study of the forward elastic proton--proton single $A_\text{N}(t)$ and double $A_\text{NN}(t)$ spin analyzing powers at two incident beam energies, 100 and 255\,GeV. The hadronic spin-flip amplitudes were reliably isolated in these measurements, and the subsequent Regge fit revealed the Pomeron contribution to both single and double spin-flip amplitudes.

Preliminary results for single-spin analyzing powers in inelastic $pp$ scattering (at low momentum transfer squared $t$ and missing mass excess $\Delta$) indicate a significant increase in the beam spin analyzing power with decreasing $\Delta$.

The intensive operation of HJET in ion beams provided $p^{\uparrow}A$ analyzing power measurements for six nuclei (deuterium, oxygen, aluminum, zirconium, ruthenium, and gold) at the 100\,GeV/nucleon, as well as energy scans for deuterium (10--100\,GeV/nucleon) and gold (3.8--100\,GeV/nucleon) beams.

Overall, the performance of the HJET in both proton and ion beams suggests its potential for proton and $^3$He beam polarimetry at the EIC with low systematic uncertainties, $\sigma_P/P~<~1\%$.

Nevertheless, to complete the HJET experimental data analysis, the following theoretical calculations and studies are still needed:
\\\noindent--\quad%
Theoretical evaluation, including normalization, of functions for Regge poles $R^\pm(s)$, Pomeron/Froissaron $P(s)$, and Odderon $O(s)$, which can be used for a Regge fit studying the energy dependence of single $r_5(s)$ and double $r_2(s)$ spin-flip $pp$ amplitudes at low $t$;
\\\noindent--\quad%
Beam and target analyzing power parametrization for inelastic $pp$ scattering (considering beam proton fragmentation and detection of the target proton) at low $t$ and $\Delta$;
\\\noindent--\quad%
Parametrization (ready to use in data analysis) for the forward $p^{\uparrow}A$ analyzing power $A_\text{N}(t,s,A)$ for $|t| < 0.02\,\text{GeV}$, $2 < A < 200$, and proton beam energy $4 < E_p < 100\,\text{GeV}$.

To decide on employing the HJET for $^3$He beam polarimetry, additional and highly detailed theoretical analyses of the proton--helion hadronic spin-flip amplitudes and the breakup corrections to the measured beam polarization may be critically important.

\section*{Acknowledgements}
  The author acknowledges the support from the Office of Nuclear Physics in the Office of Science of the U.S. Department of Energy. This work is authored by an employee of Brookhaven Science Associates, LLC under Contract No.\,DE-SC0012704 with the U.S. Department of Energy.


\providecommand{\href}[2]{#2}\begingroup\raggedright\endgroup

\end{document}